\documentclass[11pt,a4paper]{article}
\usepackage{jheppub,amsmath,amssymb,slashed,url,bm,textgreek,upgreek}
\usepackage{jheppub}  
\usepackage{tikz,lipsum,lmodern}
\usepackage[most]{tcolorbox}
\usepackage{amssymb} 
\usepackage{amsmath}
\usepackage{mathtools}
\usepackage{amsfonts}    
\usepackage{dsfont}
\usepackage{pdfpages}
\usepackage{verbatim}
\usepackage{tensor}
\usepackage{mathrsfs}
\usepackage{textgreek} 
\usepackage[mathscr]{euscript}
\usepackage[normalem]{ulem}
\usepackage{tikz}
\usetikzlibrary{decorations.pathreplacing, decorations.markings,calc,shapes.misc,decorations.pathmorphing,patterns.meta, math}

\newcommand{\beq}{\begin{equation}}
\newcommand{\eeq}{\end{equation}}
\newcommand{\nn}{\nonumber\\} 
\newcommand{\bea}{\begin{eqnarray}}
\newcommand{\ea}{\end{eqnarray}}
\newcommand{\barr}{\begin{array}}
\newcommand{\earr}{\end{array}}

\def\d{{\rm d}}
\def\i{{\rm i}}

\newcommand\AdS{\text{AdS}}
\newcommand\pin{\text{pin}}
\newcommand\T{{\sf T}}
\newcommand\F{{\sf F}}
\newcommand\PSL{\text{PSL}}

\newcommand\U{\text{U}}

\newcommand\SL{\text{SL}}
\newcommand\SU{\text{SU}}

\title{Les Houches lectures on two-dimensional gravity and holography\\
\vspace{-0.9cm}
}

 \author{Gustavo J. Turiaci}
\affiliation{Physics Department, University of Washington, Seattle, WA, USA}\emailAdd{turiaci@uw.edu}
\abstract{Lecture notes prepared for the Les Houches school ``Quantum Geometry: Mathematical Methods for Gravity, Gauge Theories and Non-Perturbative Physics” that took place during the summer 2024. We cover the techniques to perform the exact gravitational path integral of two-dimensional dilaton-gravity, and supergravity, over spacetimes with arbitrary topology, with an application to black holes. We discuss the connection with random matrix models and moduli spaces of hyperbolic surfaces briefly, since those concepts were covered in other lectures of the school.}

\begin{document}\maketitle

\thispagestyle{empty} 

\newpage
\setcounter{page}{1}
\section{Introduction}

These notes were prepared for a series of lectures presented at the 2024 Les Houches school on ``Quantum Geometry: Mathematical Methods for Gravity, Gauge Theories and Non-Perturbative Physics''. Their goal is to review the techniques used to evaluate the gravitational path integral (GPI) for theories of two-dimensional gravity, mainly Jackiw-Teitelboim (JT) gravity and its deformations, over spacetimes with nontrivial topology. This led to holographic dualities between solvable models of two-dimensional gravity and ensembles of random matrices. We will emphasize the description in terms of $BF$ theory that allows straightforward generalization beyond the simplest cases.  

\medskip 

The material and the focus taken here are complementary to that taken in the review written with T. Mertens for Living Reviews in Relativity \cite{Mertens:2022irh}, although some overlap was required to make this note self-contained. We do not discuss in depth topics such as matrix models and topological recursion, which were covered in other lectures in this Les Houches series \cite{bouchard2024leshoucheslecturenotes,JohnsonLesHouchesNotes,GiacchettoLesHouchesNotes,EynardLesHouchesNotes}. Instead, the goal here is to focus on technical aspects of the evaluation of the GPI and some generalizations that were not covered in \cite{Mertens:2022irh}, for example, including spin structures, unorientable spacetimes, and supersymmetry. The approach here is perturbative in the topological expansion, and some fundamental aspects of the non-perturbative definition of these models are covered in C. Johnson lectures \cite{JohnsonLesHouchesNotes}.

\medskip

In the rest of the introduction we give a general motivation for the type of questions we will investigate, and an outline of the lectures and topics covered.

\subsection{Motivation} 

In the 1970's it was uncovered that black holes evolve following dynamical laws that take the same form as those of thermodynamics. Perhaps the most famous proposal was that black holes have an entropy
\beq
S = \frac{A}{4 G_N},
\eeq
as well as a non-vanishing temperature. The field of black hole thermodynamics evolved over the decades and culminated in the 1990's with the advent of holography and AdS/CFT. The main lesson, whose implications we are still exploring, is that black holes not only behave as thermodynamic systems, they actually evolve as unitary quantum systems with large but finite entropy! String theory provided realizations of this relation by assigning specific quantum systems to certain black holes. This fascinating conjecture has been referred to in \cite{Almheiri:2020cfm} as the central dogma: ``As seen from the outside, a black hole can be described in terms of a quantum system with $A/4G_N$ degrees of freedom, which evolves unitarily under time evolution''.

\medskip

So far the most successful approach to studying quantum gravity is provided by the GPI, which produces answers consistent with the central dogma in highly non-trivial situations. The GPI was pioneered by Gibbons and Hawking \cite{Gibbons:1976ue} but has evolved considerably over the past several decades. It consists of formulating the experiment done on the black hole in the path integral language, using this to determine boundary conditions far from the horizon, and performing a path integral including fluctuations in the spacetime metric and topology. Both the central dogma and the interpretation of the GPI are the most precise in the context of AdS/CFT. The boundary conditions in the GPI are determined at the conformal boundary of the throat, and according to the holographic dictionary it should be reproduced by a CFT calculation.

\medskip

These lectures cover simple solvable models of quantum black holes provided by Jackiw-Teitelboim (JT) two-dimensional gravity and generalizations. This theory has gotten a lot of interest in the past decade for three main purposes:
\begin{itemize}
    \item It is a sector that describes the low-energy dynamics of strongly coupled fermionic systems such as the SYK model \cite{Chowdhury:2021qpy}.
    \item JT gravity coupled with matter captures quantum effects that become large for higher-dimensional charged black holes at low temperatures \cite{Ghosh:2019rcj,Iliesiu:2020qvm,Heydeman:2020hhw}.
    \item The black hole information paradox concerns situations where gravity seems to be in tension with the central dogma. Examples are the Page curve and the late-time behavior of correlators. Thanks to toy models such as JT gravity, it was recently discovered that spacetimes with non-trivial topologies such as spacetime wormholes play a central role in resolving these puzzles. Some of these applications of JT gravity will be the main focus of this series of lectures.
\end{itemize}

\subsection{Outline}

These notes are organized as follows. 

\medskip 

In \textbf{section \ref{sec:JTbasics}} we give a brief introduction to basic aspects of JT gravity such as its action, the physically relevant boundary conditions, and its first order formulation. We also review the derivation of the Schwarzian dynamics and its solution. This part closely follows the presentation in \cite{Mertens:2022irh}. 

In \textbf{section \ref{sec:sss}} we derive the duality between pure JT gravity and random matrix model discovered by Saad, Shenker and Stanford \cite{Saad:2019lba}. We focus mainly on the calculation of the GPI of JT gravity, particularly using the torsion, filling some gaps that were not addressed in \cite{Mertens:2022irh}. We assume a background on matrix models and its double-scaling limit since those topics were covered in Eynard \cite{EynardLesHouchesNotes} and Johnson's lectures at the school \cite{JohnsonLesHouchesNotes}. 

In \textbf{section \ref{sec:generaliz}} we generalize the duality by considering theories of pure 2d gravity with arbitrary dilaton potentials following our work with Maxfield and independently Witten \cite{Maxfield:2020ale,Witten:2020wvy}, and other generalizations studied by Stanford and Witten \cite{Stanford:2019vob} incorporating spin structure and unorientable spacetimes. 

In \textbf{section \ref{sec:sugra}} we cover the generalization of these dualities to JT supergravity, which nicely combines all the ingredients covered in the previous sections, following the original analysis of Stanford and Witten \cite{Stanford:2019vob} for $\mathcal{N}=1$, and our work with Witten  in \cite{Turiaci:2023wrh} for $\mathcal{N}=2$. In particular, we follow the $BF$ theory approach and in particular the torsion, which proved to be extremely useful in this investigation. 

In \textbf{section \ref{conclusions}} we conclude with a list of some recent research directions that we did not have time to cover.

\section{Jackiw-Teitelboim gravity -- Basics}\label{sec:JTbasics}

\subsection{Two-dimensional dilaton gravity}

We aim to construct a theory of 2d gravity with a clear semiclassical limit presenting black hole solutions that we can use to explore the consequences of the GPI. In two dimensions, the Einstein-Hilbert action itself is topological and does not suppress fluctuations
\bea
\chi &=& \frac{1}{4\pi} \Big( \int_M \sqrt{g} R + 2\oint_{\partial M} \sqrt{h} K\Big)=2-2g-n,
\ea
given by the Euler characteristic. This fact does not imply that the theory is trivial; in the path integral formulation, one needs to take care of the measure and its gauge fixing leading to an inherently strongly coupled theory. (In the context of the non-critical string this theory can be presented as the $(2,3)$ minimal model, see C. Johnson's lectures \cite{JohnsonLesHouchesNotes})

\medskip

To solve this problem, we introduce a scalar field, the dilaton $\Phi$, following \cite{Jackiw:1984je,Teitelboim:1983ux}. We consider the action, written in Euclidean signature
\beq
    I = \underbrace{-\frac{S_0}{4\pi} \Big(\int_M \sqrt{g}R+2\oint_{\partial M} \sqrt{h} K\Big)}_{\text{topological}}- \underbrace{\frac{1}{2} \int_{M} \sqrt{g}(\Phi R + U(\Phi))-\oint_{\partial M} \sqrt{h} \Phi K}_{\text{dynamical}}
\eeq
$M$ is a 2d manifold with metric $g$ and with a boundary $\partial M$ with metric $h$. The action has three terms which play different roles:

\medskip

(1) The topological term depends on the parameter $S_0$. It is responsible for suppressing topology change but does not care about perturbative metric fluctuations. (2) The dynamical term controls the classical behavior of the theory. The scalar field $\Phi$ acts as a 2d Planck ``mass''. The classical limit corresponds to regions where $\Phi$ is large, as we will see later. This term depends on a single function $U(\Phi)$, the dilaton potential. (3) Finally, the boundary term is the well-known Gibbons-Hawking-York  term that makes the variational problem well-defined \cite{Gibbons:1976ue} with Dirichlet boundary conditions, which we will motivate below\footnote{\textbf{Exercise:} It is instructive to show that this is the most general 2d dilaton gravity at the two-derivative level, up to field redefinitions and a local Weyl rescaling \cite{Banks:1990mk}.}. Some classical aspects of this theory are reviewed in \cite{Grumiller:2002nm}.

\medskip

Jackiw-Teitelboim (JT) gravity corresponds to a theory specifically with a linear dilaton potential 
\beq
U(\Phi)=-\Lambda \Phi + U_0.
\eeq
When $\Lambda \neq 0 $ we can shift $\Phi$ to eliminate $U_0$ and redefine $S_0$ to put the action in the original form. After this manipulation, the action becomes
\beq
    I_{JT} = - \frac{S_0}{4\pi} \int_{M} \sqrt{g} R- \frac{1}{2} \int_{M} \sqrt{g}\Phi (R - \Lambda)+ I_{\rm bdy}.
\eeq
The equation of motion for the dilaton imposes that classical geometries are spacetimes with constant curvature $R=\Lambda$. The equation of motion for the metric determines the spacetime profile of the dilaton. We can consider anti de Sitter (AdS) gravity with $\Lambda<0$, or de Sitter (dS)  gravity with $\Lambda>0$. When $\Lambda=0$ we cannot remove $U_0$ which remains as a physical parameter, and the theory becomes the CGHS model \cite{Callan:1992rs}. In these lectures, we will focus mainly on AdS and work in units with $\Lambda=-2$.

\medskip

One can also include matter fields. In the simplest case they do not couple to the dilaton, e.g. for a massive scalar field $\eta(x)$ we add the following term
    \beq 
    I_{\text{matter}}[g,\eta]= \frac{1}{2} \int_M \sqrt{g} \{g^{\mu\nu}\partial_\mu \eta \partial_\nu \eta + m^2 \eta^2\}.
    \eeq
When matter is included, we will assume that it takes this form. This has the advantage of making the theory solvable both classically \cite{Jensen:2016pah,Maldacena:2016upp,Engelsoy:2016xyb} and quantum mechanically \cite{Mertens:2017mtv}.

\subsection{JT gravity as a $BF$ theory}

Let us rewrite the action of JT gravity in a first-order formulation. For simplicity, consider surfaces without boundaries first.  This means that we want to replace the path integral over the metric $g_{\mu\nu}$ by the objects:
\begin{itemize}
    \item Frame one-form $e^{a}=e^a_\mu\, \d x^\mu$ with $a=1,2$. They are determined by the metric through the relation $g_{\mu\nu} = e_\mu^a e^b_\nu \delta_{ab}$. 
    \item Spin connection $\omega^{ab}=\omega^{[ab]}_\mu \, \d x^\mu$. It is not an independent field, since it is required to solve the torsion-free constraint $\d e^a + \omega^{a}_{~b} \wedge e^b = 0$.
\end{itemize}
 The dynamical bulk term in the JT gravity action can be written in terms of the frame and spin connection as
 \beq
 \frac{1}{2} \int \d^2 x \sqrt{g} \Phi (R+2) = \int_M \Phi (\d \omega + e^1 \wedge e^2) 
 \eeq
 In a quantum-mechanical treatment we need to incorporate the torsionless constraint that determines $\omega$ in terms of the frame forms. This can be corrected by integrating-in Lagrange multipliers $X^1$ and $X^2$ as follows
 \beq
\int_M \Big[ \Phi (\d \omega + e^1 \wedge e^2) + X_a (\d e^a + \omega^{a}_{~b} \wedge e^b)\Big].
 \eeq

We can now define the following quantities
\bea
A &=& e^1 \lambda^1 + e^2 \lambda^2 + \omega \lambda^3,\nn
B &=&2 \i ( X^1 \lambda^1 + X^2 \lambda^2 + \Phi \lambda^3),\nonumber
\ea
where $\{ \lambda^1,\lambda^2,\lambda^3\}$ are $2\times 2$ matrices that generate the Lie algebra $\mathfrak{sl}(2,\mathbb{R})$, chosen with the normalization condition ${\rm Tr} \,\lambda^i \lambda^j = \eta^{ij}/2$ with $\eta={\rm diag}(1,1,-1)$. The signature of $\eta$ reflects the fact that $\SL(2,\mathbb{R})$ is a non-compact group. In terms of these adjoint-valued one-form $A$ and zero-form $B$, the JT gravity action, including the torsion constraint, can be written in the suggestive form
\beq
I = -\i \int {\rm Tr}\, B F,~~~~F=\d A + A \wedge A.
\eeq
This is the action of a $BF$ theory \cite{Horowitz:1989ng} with gauge group ${\rm SL}(2,\mathbb{R})$. The path integral of this theory on a 2d surface localizes into the space $\mathcal{T}$ of flat  $\SL(2,\mathbb{R})$ connections (with $F=0$) modulo gauge transformations. The reason is that $B$ acts as a Lagrange multiplier
\beq
\int \d B\, e^{- I} = \int \d B\,\, e^{\i \int {\rm Tr} BF } = \delta (F).
\eeq
This is the origin of the factor of $\i$ in the expression for $B$, the natural contour for a Lagrange multiplier this way corresponds to real $B$, or imaginary dilaton. What is the measure of integration over moduli space? This will be addressed in the next section. Besides this choice, it is worth emphasizing that the connection between JT gravity and $BF$ theory is subtle for the following reasons:

\paragraph{Flat connection $\neq$ geometry} The moduli space $\mathcal{T}$ has multiple components distinguished by a topological invariant. Only one of these components can be related to a hyperbolic metric. (The discussion above technically also includes a choice of spin structure. To remove this, we can focus on ${\rm PSL}(2,\mathbb{R})$. We will have more discussion on such global issues later.)

\paragraph{Large Diffeomorphisms} Gravity contains large diffeomorphisms as a gauge symmetry that are not incorporated into the gauge transformations of the $BF$ description. These transformations modulo local ones are called, in the context of 2d geometry, the mapping class group. This restricts the appropriate component of flat connections $\mathcal{T}$ further to the moduli space of hyperbolic surfaces $\mathcal{M}$.

\paragraph{Sum over topologies} In gravity we should sum over topologies, as we will do in the second lecture. This is not naturally included in the gauge theory description but should be done by hand when studying gravity.

\paragraph{Boundaries} The boundary conditions natural from the gravity perspective, namely the asymptotically AdS condition, do not have any useful description in the gauge theory language. We will specify these boundary conditions more carefully below. Therefore, in the presence of boundaries, a mix of first- and second-order manipulations seems to be unavoidable. Some progress in reproducing the results in JT gravity completely from a first-order formulation are in \cite{Blommaert:2018oro,Iliesiu:2019xuh}.

\medskip

Finally we would like to point out that for general dilaton potential one can show that the gravitational action can be written locally as a Poisson sigma model \cite{Ikeda:1993aj,Ikeda:1993fh}. The form of the algebra depends explicitly on the dilaton potential, see \cite{Schaller:1994es} and \cite{Mertens:2022aou} for details.

\medskip

\subsection{Classical solutions and boundary conditions} 

Before attempting to evaluate the GPI in JT gravity we need to specify which boundary conditions we want to impose. This requires some physical considerations, which we now describe. The lack of understanding of the physically relevant boundary condition was in part the source of confusion about $\AdS_2$ in recent decades.

\medskip

First, assume that the dilaton is constant. The equation of motion that arises from varying the metric, with the assumption of a constant dilaton $\Phi$, implies that $U(\Phi)=0$ and therefore $\Phi=0$. The equation of motion for the dilaton imposes $R=-2$. This implies that locally all solutions have ${\rm AdS}_2$ metric. Globally, there can be physically different choices: 

\medskip

\paragraph{Global Patch} In Lorentzian signature, this patch has the topology of a strip with two boundaries
\beq
 \begin{tikzpicture}[baseline={([yshift=-0.1cm]current bounding box.center)}]
	 	\fill[blue!20] (-1,-1.5) rectangle (1,1.5);
   \draw[red, very thick] (-.5,-1.5) to [bend left=0] (-.5,1.5);
        \draw [black,very thick] (1,-1.5) -- (1,1.5);
        \draw [black,very thick] (-1,-1.5) to (-1,1.5);
        \node at (-1,0) [left]{\footnotesize ${\sf z}=0$};
        \node at (1,0) [right]{\footnotesize ${\sf z}=\frac{\pi}{2}$};
 \end{tikzpicture}
 \hspace{2cm}
 \d s^2 = 4\frac{ -\d {\sf t}^2 + \d {\sf z}^2}{\sin^2 2{\sf z}}
\eeq
This patch represents the maximal extension of $\AdS_2$. The red line shows the worldline of an observer sitting at a fixed spatial coordinate ${\sf z}$. The two boundaries are in causal contact, and can therefore be interpreted as an eternal wormhole connecting the two boundaries. 

\medskip

\paragraph{Poincare Patch} In Lorentzian signature this patch has a single boundary and covers a region inside the global patch
\beq
 \begin{tikzpicture}[baseline={([yshift=-0.1cm]current bounding box.center)}]
	 	\fill[blue!20] (-1,-1.7) rectangle (1,1.7);
    \draw[red, very thick] (1,-1.5) to [bend left=20] (1,1.5);
        \draw [black,very thick] (1,-1.7) -- (1,1.7);
        \draw [black,very thick] (-1,-1.7) to (-1,1.7);
        \draw[black, very thick, dashed] (1,-1.5) -- (-1,0) -- (1,1.5);
        \node at (1,0) [right]{\footnotesize $Z =0$};
         \node at (0,1.3) {\footnotesize $Z =\infty$};
 \end{tikzpicture}
 \hspace{1cm}
 \d s^2 =\frac{ -\d T^2 + \d Z^2}{Z^2}
\eeq
The red line corresponds to an observer sitting at a constant $Z$. This looks closer to a black hole with the dashed line representing the event horizon. There are regions in the bulk that are causally disconnected from the boundary observer. Nevertheless the horizon is at an infinite proper distance and has zero-temperature, making it the two-dimensional analog of vacuum AdS in higher dimensions. The metric on this patch simplifies the action of the ${\rm PSL}(2,\mathbb{R})$ isometry group
\beq
X^\pm \to \frac{a \, X^\pm + b}{c\, X^\pm + d} ,~~~~X^\pm = T \pm Z,
\eeq
where $ad-bc =1$ and the four parameters are defined up to an overall sign (which would require a fermion to be detected).

\medskip

\paragraph{Black Hole Patch} This patch truly corresponds to a black hole geometry, and is sometimes called the Rindler patch or black hole patch
\beq
 \begin{tikzpicture}[baseline={([yshift=-0.1cm]current bounding box.center)}]
	 	\fill[blue!20] (-1,-1.7) rectangle (1,1.7);
           \draw[red, very thick] (1,-1) to [bend left=20] (1,1);
         \draw[red, very thick] (-1,-1) to [bend left=-20] (-1,1);
        \draw [black,very thick] (1,-1.7) -- (1,1.7);
        \draw [black,very thick] (-1,-1.7) to (-1,1.7);
        \draw[black, very thick, dashed] (1,-1) -- (-1,1);
        \draw[black, very thick, dashed] (-1,-1) -- (1,1);
        \node at (1,0) [right]{\footnotesize $\rho =\infty$};
         \node at (0,1) {\footnotesize $\rho=0$};
 \end{tikzpicture}
 \hspace{1cm}
 \d s^2 =- \frac{4\pi^2}{\beta^2}\sinh^2\rho \, \d t^2 + \d \rho^2.
\eeq
The red lines are worldlines of observers sitting at fixed $\rho$. The dashed lines at $\rho=0$ are event horizon for observers at both boundaries located at $\rho = \infty$. There are events in the bulk that are causally disconnected, but the horizon is now at a finite distance from the boundary and has a finite temperature given by $1/\beta$. In Euclidean signature $t=- \i \tau$ and the geometry becomes the hyperbolic disk with the horizon at the origin with metric $\d s^2 = \sinh^2 \rho \,\d \theta^2 + \d \rho^2$ and $\theta = \frac{2 \pi \tau}{\beta}$ has period $2\pi$.

\medskip

According to holography, quantum gravity in $\AdS_2$ should be dual to a quantum mechanical theory with Hilbert space $\mathcal{H}$ and Hamiltonian $H$ living on the boundary. Since the black hole patch comes with two boundaries there should be two copies of that theory in some specific state. The GPI over half the disk prepares the so-called thermo-field double which is dual to \cite{Maldacena:2001kr}
\beq
|{\rm TFD} \rangle = \sum_n e^{-\beta E_n/2} |E_n \rangle_L \otimes |E_n \rangle_R,
\eeq
where $\{ E_n\}$ is the spectrum of the dual theory. The path integral over the full disk computes the overlap
\bea
\langle {\rm TFD} | {\rm TFD} \rangle &=& \int \d E ~\rho(E)~ e^{-\beta E} = {\rm Tr}_{\mathcal{H}} ( e^{-\beta H} ),~~\rho(E) = \sum_n \delta(E-E_n)\nn
&=& Z(\beta),
\ea
which can also be interpreted as the thermal partition function in the canonical ensemble of the putative quantum system. We also use this opportunity to introduce the density of states of the black hole spectrum $\rho(E)$. This interpretation is clear in Euclidean signature; the boundary of the hyperbolic disk is a circle and the path integral of a quantum system on a circle is a thermal partition function. 

\bigskip

The situation above is problematic. Which patch are we supposed to choose? What physical input determines the relation between bulk time and boundary time in the quantum mechanical description? The problem is the assumption that the dilaton is exactly constant everywhere, as we now explain.

\medskip

The solution \cite{Almheiri:2014cka,Jensen:2016pah,Maldacena:2016upp,Engelsoy:2016xyb} is to impose boundary conditions that break the conformal symmetry of $\AdS_2$ by turning on a source for the dilaton. This is something special that happens in two dimensions; in higher-dimensional versions of $\AdS/{\rm CFT}$ it is safe to work with theories that are exactly conformal invariant. To implement this, put a cutoff in the geometry and impose Dirichlet boundary conditions
    \beq\label{eq:bdycond}
\d s^2|_{\partial M}=\frac{\d \tau^2}{\varepsilon^2},~~~~~\Phi|_{\partial M}=\frac{\Phi_r}{\varepsilon},~~~\varepsilon \to 0.
    \eeq
This leads to the so-called nearly-AdS space, or ${\rm NAdS}_2$. The geometry, either in Lorentz or Euclidean signature, will now look like the following, where we draw the cut-off curve in black:
    \beq
 \begin{tikzpicture}[baseline={([yshift=-0.1cm]current bounding box.center)}]
	 	\fill[blue!20] (-1.2,-1.3) rectangle (1.2,1.3);
 \path [draw, black,very thick, decorate, decoration={random steps,segment length=3pt,amplitude=1pt}]
        (-1.3,-1.3) to[out = 80, in = -80] (-1.3,1.3);
         \draw [black,very thick, decorate, decoration={random steps,segment length=3pt,amplitude=1pt}]
        (1.3,-1.3) to[out=100, in = -100] (1.3,1.3);
        \draw [black,very thick] (1.3,-1.3) -- (1.3,1.3);
        \draw [black,very thick] (-1.3,-1.3) to (-1.3,1.3);
        \draw[black, very thick, dashed] (-1.3,-1.3) -- (1.3,1.3);
        \draw[black, very thick, dashed] (1.3,-1.3) -- (-1.3,1.3);
        \node at (0,-1.65) [below]{Lorentzian};
 \end{tikzpicture}
    \hspace{40pt}
	 \begin{tikzpicture}[scale=0.9,baseline={([yshift=-0.1cm]current bounding box.center)}]
    \filldraw[fill=blue!20, very thick, decorate, decoration={random steps,segment length=3pt,amplitude=1.5pt}] (0,0.1) circle (1.4);
 		\draw [black, very thick] (0,0.1) circle (1.55);
 		\node at (0,-1.65) [below]{Euclidean};
  \end{tikzpicture}
  \eeq
At finite temperature the black hole patch is selected, and at zero-temperatures the Poincare one. The global patch instead does not support a solution for the dilaton with the appropriate boundary conditions, unless one deforms the theory in a special way \cite{Maldacena:2018lmt}. 

\medskip 

Regardless on the fact that we turn on a source for the dilaton, the geometry can still be taken to be locally exactly AdS everywhere, and we can work in a gauge where all the information of metric fluctuations is encoded in the shape of the boundary curve consistent with \eqref{eq:bdycond}.

\medskip 

The asymptotic symmetries of $\AdS_2$ are the 1d conformal group of time reparametrizations. This group is spontaneously broken to the global conformal group $\SL(2,\mathbb{R})$ of isometries by the $\AdS_2$ background. This part is similar to the situation in 3d gravity; while the asymptotic symmetries are (two copies of) the whole Virasoro group generated by $L_n$, for $n\in \mathbb{Z}$, vacuum AdS is only invariant under the global conformal group generated by $L_{-1}, L_0$ and $L_1$. Beyond this, something special in NAdS is that the conformal symmetry is explicitly broken by the choice of boundary conditions since $\Phi_r$ is dimensionful. Another way to see this is that classically the dilaton profile is proportional to $\Phi \propto \Phi_r \cosh \rho$ and its manifestly non-invariant under the isometries of the geometry.

\paragraph{The Schwarzian mode} The off-shell action of a given boundary curve consistent with the choice of boundary conditions in \eqref{eq:bdycond} is the so-called Schwarzian action, which appeared first in the context of the SYK model \cite{Sachdev:1992fk, Sachdev:2010um,kitaevTalks, Sachdev:2015efa}. This can be done in Lorentzian signature, but since most of our calculations in the next section are carried out naturally in Euclidean signature we will choose the latter.

We work in the hyperbolic disk with coordinates $(\hat{\tau},\hat{\rho})$. The reason to relabel them is that we want to save $\tau$ for the boundary time. Let us denote the location of the boundary by $(\hat{\tau},\hat{\rho}) = (\hat{\tau}=f(\tau), \hat{\rho}=\rho(\tau))$. Since Euclidean time is compact the first variable $f(t)$ should satisfy
\beq
f(\tau) \in {\rm Diff}(S^1) ,~~~~~f(\tau+\beta) = f(\tau) + \beta.
\eeq
For a given time reparametrization $f(\tau)$ the radial coordinate $\rho(\tau)$ is determined by the Dirichlet boundary condition on the metric 
\beq
\d s^2 |_{\text{bdy}}= \Big(\rho'(\tau)^2 + \frac{4\pi^2}{\beta} \sinh^2 \rho(\tau)\,f'(\tau)^2\Big)\d \tau^2\sim  \frac{4\pi^2}{\beta} \frac{e^{2\rho}}{4} f'(\tau) \, \d \tau^2 = \frac{1}{\varepsilon^2} \d \tau^2.
\eeq
This condition determines $\rho(\tau)$ in terms of $f(\tau)$ and this relation is quite simple close to the conformal boundary of AdS since $\rho \sim - \log \varepsilon - \log f'(\tau)$. This justifies dropping the first term above.

\medskip

The boundary degree of freedom is parametrized then by a single function $f(\tau) \in {\rm Diff}(S^1)$, but this space is clearly too large. Two boundary curves related by the isometries of $\AdS_2$ should be considered equivalent even if they correspond to different profiles time reparametrizations. To find this identification, it is simpler to go to the Poincare patch coordinates, and use the fact that since we are close to the conformal boundary $Z\sim 0$. The isometry then acts as
\beq
F \to \frac{ a F + b }{ c F + d} ,~~~~F=\frac{\beta}{\pi}\tan \frac{\pi f}{\beta},~~~~ad-bc=1,
\eeq
where $F$ is the time repametrization in Poincare coordinates. Therefore the space of physically distinct boundary curves is  $\text{Diff}(S^1)/\PSL(2,\mathbb{R})$. This space is the coadjoint orbit of the Virasoro group associated to the vacuum representation, see for example \cite{Witten:1987ty}. 

\medskip

We can now evaluate the JT action on such configurations. The topological term gives $-S_0$ since $\chi = 1$ for the disk. The bulk dynamical term in JT gravity vanishes since locally $R=-2$ everywhere. The boundary term is divergent and can be regularized by a local counterterm $\oint \sqrt{h} \Phi$ which does not affect the variational problem. The action for the boundary curve, with matter sources turned off, is the Schwarzian theory\footnote{\textbf{Exercise:} Evaluate the extrinsic curvature of the boundary and reproduce this result.}
\beq
-I[f] = \underbrace{S_0}_{\text{from topological term}} +\underbrace{ \Phi_r \int_0^\beta \d \tau\, \Big\{ \tan \frac{\pi f(\tau)}{\beta},\tau\Big\}}_{\text{from dynamical terms}}
\eeq
To write the action we lifted the element of ${\rm Diff}(S^1)/\PSL(2,\mathbb{R})$ to $f(\tau)\in {\rm Diff}(S^1)$ and we consistently obtained an action which is invariant under $\PSL(2,\mathbb{R})$ representing the isometries of the disk.

\medskip 

The equation of motion of the Schwarzian action is $\frac{\d}{\d \tau} \{ \tan \pi f/\beta, \tau\}=0$. Up to a conformal transformation the solution is simply $f(\tau)=\tau$, a circle. This leads to a classical partition function $Z(\beta)$ and density of states $\rho(E)$
\beq\label{eq:actiononshelljtdisk}
\log Z \sim S_0 + \frac{2\pi^2 \Phi_r}{\beta},~~~~\rho(E) \sim e^{S_0} \, e^{2\pi \sqrt{2\Phi_r E}}.
\eeq
What happened with the conformal symmetry? Time-reparemetrizations are evidently broken by the Schwarzian action. Global conformal transformations acting on boundary time $\tau$ are also broken. If $\Phi_r$ is zero, the symmetry is unbroken but fluctuations in the boundary shape are unsuppressed \cite{Almheiri:2014cka,Jensen:2016pah,Maldacena:2016upp,Engelsoy:2016xyb}. Equivalently, the low temperature limit of JT gravity is strongly coupled since the dimensionless coupling is the temperature itself in units of $\Phi_r$. Breaking conformal symmetry is also necessary to get a reasonable spectrum; otherwise a finite-entropy spectrum could only be $\delta(E)$, with no dynamics.

\medskip

When matter is coupled to the metric but not the dilaton, the theory can also be rewritten in terms of the Schwarzian mode only even when matter sources are turned on. For example for the scalar field $\eta$ we considered earlier the effective action is 
\beq
I \to I + \int \d \tau_1 \d \tau_2\, \eta_r(\tau_1) \eta_r(\tau_2)\,\left( \frac{f'(\tau_1)f'(\tau_2)}{(\beta/\pi)^2 \sin^2 \pi (f(\tau_1)-f(\tau_2))/\beta}\right)^{\Delta} ,
\eeq
where $\eta_r(\tau)$ is a rescaled Dirichlet boundary condition for the bulk scalar field and $\Delta=1/2+\sqrt{1/4+m^2}$.  More details on this derivation are reviewed in section 3 of \cite{Mertens:2022irh}.

\subsection{Schwarzian theory -- Partition function}

Let us recall what we did, now in path-integral language. We first integrated-out the dilaton, localizing to hyperbolic metrics. For the topology of the disk there is only one choice up to a choice of boundary, reducing the integral over metrics solely to a choice of boundary curve:
\begin{align}
Z(\beta) &= e^{S_0} \int [\d g \, \d \Phi] ~ e^{ \frac{1}{2}\int_{M} \d^2 x\sqrt{g} \Phi (R+2) + \oint_{\partial M} \sqrt{h} \Phi (K-1)} \label{eqn:SchPIZJT}\\
&= e^{S_0} \int [\d f] ~e^{\Phi_r \int_0^\beta \d\tau \{ \tan \frac{\pi f(\tau)}{\beta},\tau\} }. \label{eqn:SchPIZ}
\end{align} 
The measure on the first or second line can be derived from the $BF$ analysis which comes with a natural Riemannian metric derived from a symplectic form
\beq
\Omega =2 \int {\rm Tr} \Big[ \delta_1 A \wedge \delta_2 A\Big].
\eeq
From this symplectic form one can derive a measure over the Schwarzian mode by translating the boundary curve into first-order formalism, see section 3 of \cite{Saad:2019lba}. This measure is precisely the natural one over the coadjoint orbit $\text{Diff}(S^1)/\PSL(2,\mathbb{R})$ derived  in \cite{Alekseev:1988ce,Alekseev:2020jja}.

\medskip

Multiple ways have been developed to perform the Schwarzian path integral exactly. To compute the partition function without matter sources, we can use the Duistermaat-Heckman theorem as proposed by Stanford and Witten \cite{Stanford:2017thb}. This is applicable since the integration space $X={\rm Diff}(S^1)/\PSL(2,\mathbb{R})$, being a coadjoint orbit of a group, is symplectic and the Schwarzian action acts as a Hamiltonian $H$ that generates via Poisson brackets a $\U(1)$ symmetry corresponding to time translations. In general, such integrals reduce to fixed points $P\in X$ of the $\U(1)$ symmetry
\beq
\int_X e^{\Omega}\, e^{-H} = \sum_{\text{fixed points }P } \frac{e^{-H(P)}}{\sqrt{\text{det}'D}}.
\eeq
This theorem implies two things 1) that the Schwarzian path integral is one-loop exact around fixed-points of the $\U(1)$ symmetry; and 2) that the one-loop determinant is equal to the product of ``rotation angles''. The output of the Gaussian integral is $1/\sqrt{\text{det}' D}$, where $D$ is the operator that generates the rotation symmetry of the fixed point, and the notation $\text{det}'$ means that modes of $D$ that can be generated by symmetries of the disk should be discarded. We note that the modes that are discarded are zero-modes in the sense that they do not appear in the action or in the symplectic form, but they are in general not zero-modes of $D$. So in general a few eigenvalues of $D$ are omitted by hand.

\medskip

In our case we have $D=\partial_\tau -  \partial_f$. The second term is required so that $f(\tau)=\tau$ is indeed annihilated by $D$. A Fourier mode expansion of $\delta f = f-\tau$ provides a basis of eigenfunctions of $D$, $\delta f \sim e^{ 2\pi\i n \tau/\beta}$ with $n \in \mathbb{Z}$. The linearization of the isometries leads to fluctuations with $n=-1,0,1$ and hence restricts the range over $n$. The rotation angles are then $\Phi_r n/\beta$, with a normalization determined by the normalization of the symplectic form, see \cite{Saad:2019lba}. The final answer is
\begin{equation}
    Z(\beta) ~~=~~ e^{S_0 + \frac{2\pi^2\Phi_r}{\beta}} \prod_{n\geq 2} \frac{\beta}{\Phi_r n} 
    ~~=~~ \frac{\Phi_r^{3/2}}{4\sqrt{\pi} \beta^{3/2}} e^{S_0 + \frac{2\pi^2\Phi_r}{\beta}}
    .\label{eq:oneloopZSCH}
\end{equation}
   The infinite product was taken via zeta-function regularization. We see that if we turn off the conformal symmetry-breaking deformation $\Phi_r \to 0 $ the one-loop determinant vanishes. This means that there is no natural definition of a regulated volume of ${\rm Diff}(S^1)/\SL(2,\mathbb{R})$. Since $\Phi_r$ is dimensionful, from now on we will work in units with $\Phi_r=1/2$ for simplicity. In these units, the coupling constant is the temperature/energy. 

\medskip

The central dogma implies that one should be able to interpret \eqref{eq:oneloopZSCH} as the free energy of a quantum system. We can use the exact  Schwarzian partition function to infer what the density of states of this quantum system should be. An inverse Laplace transform of Eq.~\eqref{eq:oneloopZSCH} gives
\begin{equation}\label{eqn:JTexactRho}
\rho_{\rm JT}(E) = \frac{e^{S_0}}{4\pi^2} \sinh \left( 2\pi \sqrt{E} \right).
\end{equation}
This result is nonperturbative in the Schwarzian coupling, which suppresses perturbative metric fluctuations, but only leading order in $S_0$. Surprisingly, JT gravity on the disk with matter is also exactly solvable, even though correlators are not one-loop exact and no localization argument applies. The path integral with matter sources can be derived using the relation between ${\rm Diff}(S^1)/\PSL(2,\mathbb{R})$, representation theory of Virasoro, and Liouville CFT. This was the approach we had with Mertens and Verlinde \cite{Mertens:2017mtv}. The results, and an account of different ways to derive them such as the particle in-the-magnetic-field approach of Kitaev, Suh and Yang \cite{Kitaev:2018wpr,Yang:2018gdb}, are reviewed in section 3 of \cite{Mertens:2022irh}.

\subsection{The wormhole length}

In this section we review a different perspective on the disk partition function that, although we will not need it in the rest of the lectures, it is quite useful for some applications. 

\medskip 

We saw above that the GPI on the half-disk prepares the thermofield double state $|{\rm TFD}\rangle$ on a pair of entangled black holes. We can think of the bulk gravity states as being labeled by the (renormalized) length of the wormhole $\ell$. This parameter takes a temperature-dependent value on-shell but can present quantum fluctuations. Therefore, in gravity the wave function of the state can be described by a function of the wormhole length $$\Psi_{\beta}(\ell)= \langle \ell | {\rm TFD}_\beta \rangle,$$ where $\beta$ parameterizes the amount of (Euclidean) proper time evolution on the boundary. The states here refer to the bulk Hilbert space, not the boundary one.
\beq
	 \begin{tikzpicture}[scale=0.9,baseline={([yshift=-0.1cm]current bounding box.center)}]
    \filldraw[fill=blue!20, very thick, decorate, decoration={random steps,segment length=3pt,amplitude=1.5pt}] (0,0.1) circle (1.4);
 		\draw [black, very thick] (0,0.1) circle (1.55);
        \draw[very thick] (-1.4,0) to [in=180+20, out=-20] (1.4,0);
        \node at (0,0) {$\ell$};
 		\node at (0,-1.65) [below]{$\beta/2$};
  \end{tikzpicture}
  \eeq
In JT gravity we can compute the wavefunction exactly using canonical quantization, combined with some input from the GPI. This was spelled out in \cite{Kitaev:2018wpr,Yang:2018gdb,Harlow:2018tqv} and more recently in \cite{Lin:2022rzw,Lin:2022zxd}. The Wheeler-de Witt Hamiltonian acting on fixed-length wavefunction, as derived from the JT gravity action, reduces to 
$$
H = - \partial_{\ell}^2 + e^{-\ell}.
$$
This Hamiltonian was also obtained earlier in \cite{Bagrets:2016cdf} and \cite{Mertens:2017mtv} using a different perspective. The eigenfunctions of this Hamiltonian are, up to normalization, $ K_{2 \i s}(2 e^{-\ell/2})$ where $E_s=s^2$ is the eigenvalue of the Hamiltonian which is the energy. One can argue that the Hartle-Hawking state at temperature $\beta$ can be expanded as 
\beq
\Psi_\beta (\ell) = \int_{-\infty}^\infty \d s\, \mathcal{A}_s\, K_{2 \i s} (2 e^{-\ell/2}),
\eeq
where $\mathcal{A}_s$ is an amplitude that we need to determine. As far as I know, this amplitude was not determined purely from a bulk canonical quantization of gravity, instead by comparing with the disk density of states computed via the GPI (either through localization, through the connection to Liouville, or the particle-in-a-magnetic field approach). The answer is
\beq
\mathcal{A}_s = e^{S_0/2}e^{-\beta E_s/2} \frac{ s \sinh(2\pi s)}{\pi^2},
\eeq
which guarantees $\int_{-\infty}^{\infty} \d \ell\,  \Psi_\beta (\ell) \Psi_\beta(\ell) = Z(\beta)$. This gives a useful representation of the matter boundary correlators computed in \cite{Mertens:2017mtv} since for example 
\bea
{\rm Tr} \, [e^{-u' H} O^\dagger e^{-u H} O] &=& \langle {\rm TFD}_{u'} | e^{-\Delta \ell} | {\rm TFD}_u \rangle,\nonumber\\
&=& \int_0^\infty \d E \rho(E) \int_0^\infty \d E' \rho(E') e^{-u E- u' E'} \frac{\Gamma(\Delta \pm \i \sqrt{E}\pm \sqrt{E'})}{  \Gamma(2\Delta)}
\ea
interpreted as the inner product between two preparations of the TFD with a wormhole of renormalized length $\ell$, with the insertion of the matter propagator $e^{-\Delta \ell}$. More details and generalizations of this result can be found in section 3 of the LRR review \cite{Mertens:2022irh}. This interpretation was exploited in \cite{Saad:2019pqd} as well as in \cite{Lin:2022rzw,Lin:2022zxd} to generalize these results to $\mathcal{N}=2$ JT gravity with interesting applications to BPS chaos.

\subsection{JT gravity and near-extremal black holes}

Besides being a toy model of quantum gravity (and its relevance to the SYK model which we will not have time to cover) JT gravity also describes the dynamics of certain higher-dimensional geometries. Near-extremal black holes universally have an $\text{AdS}_2\times X^{D-2}$ throat with an emergent isometry that includes the 1d conformal group. We can consider this in an asymptotically $M^D$ space, which might be AdS or flat. 
It is useful to study the dynamics separately in the throat, and in the far-away region, and glue their separate contributions to the path integral\footnote{The quantum effects mentioned below were recently reproduced directly from the full geometry without the need for such gluing procedure \cite{Kolanowski:2024zrq}.}. This is an old idea implemented in multiple examples; some recent references are \cite{Nayak:2018qej,Moitra:2018jqs,Castro:2018ffi,Sachdev:2019bjn}. 

For example, in the case of the Reissner-Nordstrom in four dimensions:
 \beq
 \hspace{-1cm}
\begin{tikzpicture}[scale=1.7,baseline={([yshift=0cm]current bounding box.center)}]
\draw[very thick,red] (.5,-.5) to[bend right =10] (.4,-0.9);
\draw[very thick,red] (-.5,-.5) to[bend left =10] (-.4,-0.9);
\draw[very thick,red] (.5,.5) to[bend left =10] (.4,0.9);
\draw[very thick,red] (-.5,.5) to[bend right =10] (-.4,0.9);
\draw[very thick,blue,decorate, decoration={random steps,segment length=3pt,amplitude=0.5pt}] (-.5,-.5) to[bend right =15] (-.5,.5);
\draw[very thick,blue,decorate, decoration={random steps,segment length=3pt,amplitude=0.5pt}] (.5,-.5) to[bend left =15] (.5,.5);
\fill[blue!50,nearly transparent]    (-.4,-0.9) to[bend right =10] (-.5,-.5) to[bend right =15] (-.5,.5) to[bend right =10] (-.4,0.9) -- (.4,0.9) to[bend right =10] (.5,.5) to[bend right =15] (.5,-.5) to[bend right =10] (.4,-0.9) -- (-.5,-0.9);
\draw[very thick] (-.5,-.5) -- (-1,0) -- (-.5,.5); 
\draw[very thick] (.5,-.5) -- (1,0) -- (.5,.5); 
\draw[very thick] (-.5,.5)--(.5,-.5);
\draw[very thick] (.5,.5)--(-.5,-.5);
\draw[very thick,decoration = {zigzag,segment length = 1.5mm, amplitude = 0.3mm},decorate] (-.5,.5)--(-.5,0.9);
\draw[very thick,decoration = {zigzag,segment length = 1.5mm, amplitude = 0.3mm},decorate] (.5,.5)--(.5,0.9);
\draw[very thick,decoration = {zigzag,segment length = 1.5mm, amplitude = 0.3mm},decorate] (-.5,-.5)--(-.5,-0.9);
\draw[very thick,decoration = {zigzag,segment length = 1.5mm, amplitude = 0.3mm},decorate] (.5,-.5)--(.5,-0.9);
\draw[above] (0,.5) node {\footnotesize ${\rm AdS}_2\times S^2$};
\end{tikzpicture}
\hspace{1cm}\rightarrow \hspace{1.2cm}
\begin{tikzpicture}[scale=1.7,baseline={([yshift=0cm]current bounding box.center)}]
\draw[very thick,red] (.5,-.5) to[bend right =10] (.4,-0.9);
\draw[very thick,red] (-.5,-.5) to[bend left =10] (-.4,-0.9);
\draw[very thick,red] (.5,.5) to[bend left =10] (.4,0.9);
\draw[very thick,red] (-.5,.5) to[bend right =10] (-.4,0.9);
\draw[very thick,blue,decorate, decoration={random steps,segment length=3pt,amplitude=0.5pt}] (-.5,-.5) to[bend right =15] (-.5,.5);
\draw[very thick,blue,decorate, decoration={random steps,segment length=3pt,amplitude=0.5pt}] (.5,-.5) to[bend left =15] (.5,.5);
\fill[blue!50,nearly transparent]    (-.4,-0.9) to[bend right =10] (-.5,-.5) to[bend right =15] (-.5,.5) to[bend right =10] (-.4,0.9) -- (.4,0.9) to[bend right =10] (.5,.5) to[bend right =15] (.5,-.5) to[bend right =10] (.4,-0.9) -- (-.5,-0.9);
\draw[very thick] (-.5,.5)--(.5,-.5);
\draw[very thick] (.5,.5)--(-.5,-.5);
\draw[very thick] (-.5,-0.9)--(-.5,0.9);
\draw[very thick] (.5,-0.9)--(.5,0.9);
\draw[above] (0,.5) node {\footnotesize ${\rm AdS}_2$};
\end{tikzpicture}
\eeq
The gluing to the asymptotically flat region (which breaks conformal symmetry) selects the Rindler patch time as the physical one. Furthermore, asymptotic observable determines boundary condition at the throat. For example, the entropy arises mainly from near-horizon; Hawking radiation spectrum is determined by the ${\rm AdS}_2$ boundary two-point function. 

\medskip 
The higher-dimensional gravity theory in the throat is equivalent to JT gravity coupled to matter. Some comments on its derivation:

\vspace{-2pt}
\paragraph{JT-gravity sector} $S_0$ is the extremal Bekenstein-Hawking entropy. The JT metric $g_{\mu\nu}$ arises from the 4d metric along the temporal and radial directions. The dilaton $\Phi$ measures the deviation of $\text{Area}(X)$ from the extremal value. Near the horizon $|\Phi|\ll S_0$, implying non-linear dilaton-potential terms suppressed by powers of $1/S_0$.

\paragraph{Matter sector} Arises from all other KK modes of higher-dimensional metric and matter. The radius of curvature of AdS and $X$ are comparable, implying a large number of light matter fields. 2d gauge fields arise from higher-D gauge fields or isometries of $X$. Near the horizon $|\Phi|\ll S_0$, which implies that matter and dilaton interactions are suppressed by powers of $1/S_0$.

\paragraph{Boundary Condition} By including the main corrections from $\AdS_2$ induced by the gluing to the far-away region one can derive the ${\rm NAdS}_2$ boundary conditions \eqref{eq:bdycond} and extract $\Phi_r$, e.g. \cite{Almheiri:2016fws}. 

\medskip 

In the large $S_0$ limit, which can be achieved by a macroscopic black hole and small $G_N$, we can ignore both the topology change and interactions between matter with itself or the dilaton. This means that for large $S_0$ but arbitrary $\Phi_r/\beta$ we can import the above results. The quantum effects from the Schwarzian theory that we discussed earlier resolved some long-standing puzzles regarding black hole thermodynamics \cite{Preskill:1991tb,Maldacena:1998uz,Page:2000dk}, as shown in \cite{Ghosh:2019rcj,Iliesiu:2020qvm,Heydeman:2020hhw,Rakic:2023vhv,Kapec:2023ruw}. This would be a topic of a separate set of lectures, for a short summary see \cite{Turiaci:2023wrh}.

\section{Sum over topologies and random matrices}\label{sec:sss}

\subsection{Discreteness from gravity}
We have seen that JT gravity has two couplings: $\Phi_r/\beta$, suppressing perturbative fluctuations, and $S_0$, suppressing topology. This is again something special about two dimensions since in higher dimensions both roles are played by $G_N$. We have solved the theory exactly in the first coupling. What about the second? 

\medskip

The exact solution in $\Phi_r$ displays a continuum spectrum, contrary to the holographic expectation that the dual Hilbert space is discrete and has a finite entropy captured by the Bekenstein-Hawking formula. This issue is related to the information paradox stated by Maldacena \cite{Maldacena:2000dr} that we now recall. Take the two-point function of some matter field and evaluate it at late times. For a quantum system with a discrete spectrum we have, for early times,
\beq
\frac{1}{Z}{\rm Tr} \left[ e^{-\beta H} O(t) O(0)\right] = \sum_{n,m} e^{-\beta E_n} e^{-\i t(E_m-E_n)} | \langle n |  O | m \rangle |^2 \sim \text{Order one}.
\eeq
The problem arises because if we compute this quantity in gravity it decays exponentially with time for arbitrarily late times. The rates of decay are referred to as the black hole quasinormal mode spectrum. For a discrete spectrum the late-time behavior above implies the correlator should not be too small. On average the dominant contribution should come from configurations with $E_n = E_m$ such that at late times
\beq
\frac{1}{Z}{\rm Tr} \left[ e^{-\beta H} O(t) O(0)\right] \sim \sum_n e^{-\beta E_n} | \langle n | O | n \rangle |^2 \sim \text{Order }e^{-S_0}.
\eeq
Therefore the problem is correlated to the expectation that the black hole quantum system is finite dimensional so that $e^{S_0} < \infty$. In fact all versions of the black hole information paradoxes are cored in the tension between a discrete spectrum and gravity. 

\medskip

It is useful to simplify the problem even further. We can define an observable that is independent of a choice of operator and moreover is applicable even in pure gravity, where there is no such notion of a preferred operator in the bulk other than the Hamiltonian itself. This is the \textbf{spectral form factor} 
\beq
{\rm SFF}(t) = \sum_{n,m} e^{-(\beta+\i t) E_n} e^{-(\beta - \i t) E_m},
\eeq
a product of partition functions $Z(\beta_1)Z(\beta_2)$ with $\beta_{1/2} = \beta \pm \i t$. This was introduced in the context of gravity in \cite{Cotler:2016fpe}. The quantity starts off at $t=0$ as order $e^{2S_0}$ and oscillates erratically around the late-time mean $Z(2\beta)$ which is order $e^{S_0}$ and therefore suppressed by a factor of $e^{-S_0}$ with respect to early times. Its shape is roughly as (figure taken from \cite{Mertens:2022irh})
$$
\includegraphics[scale=0.5]{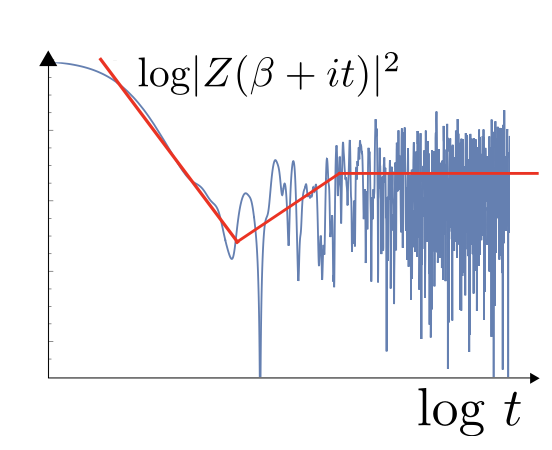}
$$
The blue line is a member drawn from the GUE ensemble, while the red line arises from averaging over Hamiltonians. One could also average over time windows for a single Hamiltonian. The ramp in the curve is characteristic of level repulsion.
\medskip

Saad, Shenker and Stanford \cite{Saad:2019lba} proposed to study pure JT gravity to explore a resolution of the black hole paradox explained above. They show that it is necessary to include spacetime wormholes when working at finite $S_0$. In particular they show pure JT gravity is equal to an average over quantum mechanical theories with a discrete spectra
\beq
Z_{\rm gravity}(\beta_1,\ldots ,\beta_n) = \int \d H \, P(H)\, {\rm Tr} \, e^{-\beta_1 H}\ldots {\rm Tr} \, e^{-\beta_n H}.
\eeq
In this context the average level spacing is of order $e^{-S_0}\ll 1$. Pure gravity captures the average part of the spectral form factor. The precise ensemble of theories will involve a matrix potential and a double-scaling limit such that the spectral curve, defined through $y(x\pm \i \epsilon) = \mp\i \pi e^{-S_0}\rho_{\rm disk}(x)$ to leading order in the large $e^{S_0}$ limit, is 
\beq
\rho_{\rm disk} (E) = \frac{e^{S_0}}{4\pi^2 } \sinh (2\pi \sqrt{E}),~~~\Rightarrow~~~y(x) = \frac{1}{4\pi} \sin (2\pi \sqrt{-x}).
\eeq
Matrix models were covered in the first week of the school and from Johnson's lectures, so some familiarity with this language will be assumed. To be more precise the GPI of JT gravity on connected spacetimes can be expanded at large $e^{S_0}$ as
\beq
Z_{\text{gravity, conn}}(\beta_1,\ldots,\beta_n) \sum_{g} \frac{Z_{g,n}(\beta_1,\ldots,\beta_n)}{(e^{S_0})^{2g+n-2}}
\eeq
We will show that order by order, this expansion matches the 't Hooft expansion of a random matrix in the double-scaling limit. 

\medskip

A non-perturbative completion of this duality in $e^{-S_0}$ is covered in C. Johnson lectures \cite{JohnsonLesHouchesNotes}, based on the original references \cite{Johnson:2019eik,Johnson:2020exp,Johnson:2021zuo}. Such an understanding is needed for seeing generic features of the late-time plateau, and seeing more explicitly the emergence of the discrete spectra of members of the ensemble. Neither of these are directly accessible using the perturbative approach. 

\medskip

We are not proposing that all holographic duals of any black hole should involve disorder. We do expect the dual to have a chaotic spectrum. This is defined as a spectrum that shares statistical features with a random matrix, without necessarily being one. JT gravity is then simply a toy model that isolates the features of gravity responsible for chaos, namely spacetime wormholes.

\subsection{Two-boundary wormhole}\label{sec:2bdywh}

We shall begin by considering the two-boundary wormhole with $g=0$ and $n=2$. This will be part of the building block for the general answer later. 

\medskip

The JT gravity path integral localizes to an integral over moduli space of hyperbolic surfaces with no handles and two boundaries. If we ignore boundary modes (which will be incorporated later), it seems that the only modulus is the length of the interior geodesic, together with a twist, which we will characterize below. How can we recognize the moduli from the point of view of $BF$ theory? A flat connection can be described by its holonomies around its non-trivial cycles. In this case we have only one such cycle, which we can choose to be the interior geodesic. We denote the holonomy by $U\in G$. Two holonomies related by conjugation $U \to R U R^{-1}$ with $R\in G$ are considered gauge-equivalent so we only care about the conjugacy class. 

\medskip

For a flat connection to be associated with a smooth geometry, the holonomy should be hyperbolic. Any hyperbolic element $U$ can be conjugated to
\bea\label{eq:Ul}
U  = \pm \left(\begin{array}{cc}
 e^{b/2} & 0 \\
0& e^{-b/2}
\end{array}\right),
\ea
where $b$ is the length of the geodesic. Given any presentation of $U$ the length can be extracted from its trace ${\rm Tr}\, U = \pm 2 \cosh(b/2)$. If we work with bosonic JT gravity, the overall sign can be discarded since we are working with $\PSL(2,\mathbb{R})$. In the presence of a spin structure, the overall sign indicates whether fermions are antiperiodic (NS) or periodic (R) around such cycle. This will be important in later lectures.

\medskip

The length of the geodesic is not the only moduli. In the $BF$ perspective on JT gravity, gauge transformations are constrained to be trivial along the boundaries. We can define a gauge-invariant ``holonomy'' $V$ by parallel transport from one boundary to the other. One can show that the following combination $V U  V^{-1} U^{-1}$ represents the holonomy of a contractible cycle which should therefore be trivial. This implies that $V$ must commute with $U$ when the connection is flat, so they must be diagonal on the same basis $V = \pm \text{diag}(e^{\varrho/2},e^{-\varrho/2})$. To avoid overcounting we restrict $0\leq \varrho \leq b$. The interior moduli are therefore the length $b$ and twist $\varrho$.

\medskip

It is useful to give a geometric meaning of the twist. We can separate the hyperbolic cylinder into two trumpets internally bounded by the geodesic of length $b$. However, there are new moduli that arise from the gluing because we can act with a global rotation before gluing. This gluing is precisely represented in the $BF$ description through $V$. 
\vspace{5pt}
\beq
\begin{tikzpicture}[scale=0.7, baseline={([yshift=+0.1cm]current bounding box.center)}]
\node at (-2.8,0) {\small $\beta_1$};
\node at (2.8,0) {\small $\beta_2$};
\draw[thick] (-2,0) ellipse (0.3 and 1.5);
\draw[thick,dashed] (0,0) ellipse (0.1 and 0.6);
\draw[thick] (2,0) ellipse (0.3 and 1.5);
\draw[thick,red] (-1,0) -- (0.1,0) to [bend right=10] (0.085,0.4) -- (1,0.4) ;
\draw[thick,red,->] (-0.50,0) -- (-0.45,0);
\draw[thick,red,->] (0.50,0.4) -- (0.55,0.4);
\draw[thick] (-1.937,1.47) to [bend right=50] (1.937,1.47);
\draw[thick] (-1.937,-1.47) to [bend left=50] (1.937,-1.47);
\node at (0,-1.985) {};
\node at (0.6,0) {\textcolor{red}{\small $\varrho$}};
\node at (0,-1) {\small $b$};
\end{tikzpicture}
\eeq
This picture hopefully clarifies why $\varrho \sim \varrho + b$. This is perhaps the simplest instance where the moduli of flat connections in the hyperbolic component (which would not put any constraint on $\varrho$) is distinguished from moduli space of hyperbolic surfaces.

\medskip
What is the interpretation of the overall sign of $V$ in a theory with fermions? Before we glue the two trumpets we can change the sign of fermions, producing a non-equivalent spin structure. For a given choice of NS/R boundaries, the cylinder will therefore have two choices of internal spin structure.
\medskip

The symplectic measure over length and twist is $\d b \, \d \varrho$, as we will see later. Since the path integral over the trumpets naturally do not depend on the twist parameter which is inherently associated to the gluing, we can integrate-out $\varrho$ from the beginning, leading to an effective measure $b \, \d b$ over geodesic lengths.
\medskip

Having described the interior moduli, we need to evaluate the path integral over the boundary modes at each NAdS boundary described by the Schwarzian theory. We can then glue all contributions as shown in the figure
\beq
\begin{tikzpicture}[scale=0.6, baseline={([yshift=0.1cm]current bounding box.center)}]
\node at (-2.8,0) {\small $\beta_1$};
\node at (2.8,0) {\small $\beta_2$};
\draw[thick] (-2,0) ellipse (0.3 and 1.5);
\draw[thick] (2,0) ellipse (0.3 and 1.5);
\draw[thick] (-1.937,1.47) to [bend right=50] (1.937,1.47);
\draw[thick] (-1.937,-1.47) to [bend left=50] (1.937,-1.47);
\node at (0,-1.985) {};
\end{tikzpicture}
~~=\int_0^\infty  \d b \, \int_0^b \d \varrho~~~~
\begin{tikzpicture}[scale=0.6, baseline={([yshift=-0.1cm]current bounding box.center)}]
\draw[thick] (6.5,0) ellipse (0.3 and 1.5);
\draw[thick] (8.5,0) ellipse (0.1 and 0.7);
\draw[thick] (6.54,1.49) to [bend right=20] (8.5,0.7);
\draw[thick] (6.54,-1.49) to [bend left=20] (8.5,-0.7);
\draw[thick] (11.5,0) ellipse (0.3 and 1.5);
\draw[thick] (9.5,0) ellipse (0.1 and 0.7);
\draw[thick] (11.46,1.49) to [bend left=20] (9.5,0.7);
\draw[thick] (11.46,-1.49) to [bend right=20] (9.5,-0.7);
\draw[thick,red,->] (9.1,0) to [bend right=20] (9,0.5)  ;
\node at (9,0.8) {\textcolor{red}{\small $\varrho$}};
\node at (8.5,-1) {\small $b$};
\node at (9.5,-1) {\small $b$};
\end{tikzpicture}
\eeq
The calculation is actually very similar to that on the disk. The path integral localizes into hyperbolic surfaces and therefore there is no bulk contribution. Since the interior boundary is a geodesic $K=0$ and the boundary term vanishes there. We end up with a theory very similar to the Schwarzian appearing on the disk but with a slightly different action depending on $b$. The integration manifold is now
\beq
\text{Diff}(S^1)/\U(1),
\eeq
since the presence of the inner boundary breaks the group of isometries of the hyperbolic disk into only rotations. The rotation angles are nevertheless insensitive to the parameter $b$ other than the fact that fewer modes are removed by isometries. We instead get a product of rotation angles
\beq
\prod_{n\geq 1} \frac{2 \beta}{n} = \frac{1}{\sqrt{4\pi \beta}}.
\eeq
Combining this with the value of the Schwarzian action evaluated on the fixed-point gives
\beq
Z_{\rm JT}^{\rm trumpet}(\beta,b) = \frac{1}{\sqrt{4 \pi \beta}} \, e^{-\frac{b^2}{4\beta}}.
\eeq
Since the trumpet has the topology of an annulus, $S_0$ does not appear here. We can now assemble the pieces and glue the contributions to the path integral from both trumpets together with the measure over interior moduli
\bea
Z_{0,2}(\beta_1,\beta_2) &=& \int_0^\infty b \, \d b \,\, Z_{\rm JT}^{\rm trumpet} (\beta_1, b)\,Z_{\rm JT}^{\rm trumpet} (\beta_1, b),\\
&=& \frac{1}{2\pi} \, \frac{\sqrt{\beta_1\beta_2}}{\beta_1 + \beta_2}.
\ea
Some comments on this result:

\medskip

\paragraph{Match with matrix integral} This is a very robust evidence in favor of the duality between JT gravity and random matrix models. In the double-scaling limit the leading order connected average of a product of two partition functions is universal (in a given symmetry class):
\beq
\Big\langle {\rm Tr}\, e^{-\beta_1 H} \,\,{\rm Tr}\, e^{-\beta_2 H} \Big\rangle_{g=0,n=2}^{\rm conn.} = \frac{1}{2\pi} \, \frac{\sqrt{\beta_1\beta_2}}{\beta_1 + \beta_2} \, e^{-\beta E_0}.
\eeq
This matches our result in JT gravity since we are working in units where the threshold energy vanishes $E_0=0$. 

\paragraph{The ramp} We can analytically continue this result into complex boundary lengths $\beta_1 = \beta/2 + \i T$ and $\beta_2 = \beta/2 - \i T$, producing precisely the spectral form factor. The two-boundary wormhole will therefore produce precisely the ramp
\beq
\frac{1}{2\pi} \, \frac{\sqrt{\beta_1\beta_2}}{\beta_1 + \beta_2} = \frac{1}{2\pi \beta} \sqrt{ \frac{\beta^2}{4} + T^2}  \to \frac{1}{2\pi} \frac{T}{\beta}.
\eeq
Upon inverse Laplace transform this is a direct consequence of level repulsion, which in the density correlator leads to $\langle \rho(E_1) \rho(E_2)\rangle_{\rm conn} \sim - (E_1-E_2)^{-2}$.

\paragraph{Three-boundary wormhole} We can easily extend this calculation to a surfaces with no handles and three boundaries. We can glue now three trumpets into a hyperbolic three-holed sphere. The simplification in this case arises because there is a single hyperbolic three-holed sphere with given boundary lengths $b_1, b_2$ and $b_3$. The gluing measure we derived applies independently to each of the three boundaries giving
\beq
Z_{0,3}(\beta_1,\beta_2,\beta_3) = \int_0^\infty \prod_{i=1,2,3} b_i \, \d b_i \, Z_{\rm JT}^{\rm trumpet}(\beta_i,b_i) = \frac{\sqrt{\beta_1\beta_2\beta_3}}{\pi^{3/2}}.
\eeq
\textbf{Exercise} show that this is the result predicted by a matrix model with the loop equations corresponding to spectral curve $y(x) = \frac{1}{4\pi} \sin (2\pi \sqrt{-x})$. 

\subsection{One-loop determinants and the torsion} 

Surface with more boundaries or more handles will inevitably come with internal moduli that need to be integrated over. All NAdS boundaries of a hyperbolic surface can be connected to a geodesic via a trumpet and therefore we can focus first on hyperbolic surfaces with geodesic boundaries. In this section, following Stanford and Witten \cite{Stanford:2019vob}, we will explain how to obtain the measure of integration over the moduli using torsion. 

\medskip

To define the path integral of a gauge theory with connection $A$ one starts with a Riemannian metric on the fluctuation field $\delta A$ such as $| \delta A|^2 = \int \text{Tr}\,\, \delta A \wedge \star \delta A$, which induces a Riemannian measure over the space of connections. In $BF$ theory we only need to integrate over flat connections $A_0$ such that $\d A_0 + A_0 \wedge A_0=0$.  The Riemannian measure above reduces to a measure, which we call $\mu_0$, on the space of zero-modes around $A_0$ that preserve flatness. A one-loop calculation in the framework of the Fadeev-Popov formalism leads to the following quantum-corrected measure over the moduli space of zero-modes 
\beq\label{eq:muqc}
\mu = \mu_0 \, \cdot \, \frac{{\rm det}'\Delta_0}{\sqrt{{\rm det}'\Delta_1}}=\mu_0 \, \cdot \, \frac{\sqrt{{\rm det}'\Delta_0}}{\sqrt{{\rm det}'\Delta_2}}.
\eeq
We expand around a given flat connection $A_0$ and define $D = \d + [A_0,\cdot]$ that maps adjoint-valued $q$-forms to $(q+1)$-forms. The Laplacian acting on adjoint-valued forms that appears in the measure is $\Delta = D^* D + D D^*$. The first equality in \eqref{eq:muqc} makes it clear that the numerator comes from the ghosts and the denominator from the gauge field. A derivation can be found in section 2.2 of \cite{Witten:1991we}. As explained in \cite{Stanford:2019vob}, the second identity arises from the Hodge decomposition of forms which implies that $\text{det}'\Delta_1 = \text{det}'\Delta_0 \,\text{det}'\Delta_2 $. On orientable manifolds we can define a Hodge star operator implying that $\text{det}'\Delta_2 = \text{det}'\Delta_0$ and therefore JT gravity is ``tree-level exact'' at the level of the measure. This simplification is not available for unorientable manifolds.

\medskip

In any dimension, the analytic torsion was defined by Ray and Singer as a certain ratio of determinants times a classical measure $\mu_0$ on the space of zero-modes \cite{RaySinger} and later shown by Schwarz to be related to gauge theories \cite{Schwarz:1978cn}. In two dimensions, the analytic torsion is precisely reduced to the measure $\mu$ above. The important result is that analytic torsion is equivalent to the combinatorial torsion studied earlier by Reidemeister \cite{Reidemeister}. This is a quantity that can be evaluated on a lattice and the result is completely independent of the lattice. In the continuum limit, the definition reduces to the analytic torsion but we can instead choose the simplest triangulation for its explicit evaluation. The connection between analytic and combinatorial torsion was proven by Cheeger and Muller for compact groups and later generalized to non-compact groups \cite{Cheeger,Muller1,Muller2,BismutZhang}.

\medskip

In conclusion we have a few possible methods of calculation. (I) For orientable surfaces the measure reduces to the bare Riemannian one $\mu_0$ which could be evaluated from first principles. (II) The moduli space of flat connections on orientable surfaces is symplectic, and a measure can be derived from the symplectic form, leading to the Weil-Petersson measure that appeared in other lectures in the school. (III) We can evaluate the measure through the combinatorial torsion on the simplest possible triangulation. 

Approach I has not lead to any useful computation and is not applicable for unorientable surfaces, since one would have to evaluate the analytic torsion anyways. Approach II is useful on orientable surfaces and leads to the Weil-Petersson measure, but has the drawback of not being applicable for unorientable surfaces or not being practical for more complicated generalizations such as supersymmetry. We will follow approach III since the combinatorial torsion is relatively easy to compute and has the advantage of being straightforward to generalize to unorientable surfaces as well as supersymmetry.

\medskip

Before introducing the combinatorial torsion in more detail, we would like to mention a subtlety that will be important later. The geodesic boundary condition is not obviously related to the boundary condition implicit in the torsion calculation we review below. If $\Sigma$ is an oriented two-manifold of genus $g$ with $n$ boundary circles, then the symplectic structure determines a measure $\mu$ on what we will call $\mathcal{T}_{g,n}$, the moduli space of flat bundles with
prescribed conjugacy classes of the holonomies 
around the boundaries; the torsion instead defines a measure $\tau$ on the moduli space of flat bundles
over $\Sigma$ without a restriction on the boundary holonomies. 
We will determine the relation between these two objects after evaluating the torsion in some simple cases.

\subsubsection*{The combinatorial torsion}

 A precise definition of the torsion can be found in section 3.2 of \cite{Stanford:2019vob} or section 4.2 of \cite{Witten:1991we}. For completeness we will summarize here some important properties to make these notes as self-contained as possible.

 \medskip 
 
 The combinatorial torsion can be thought of as being formulated in a framework dual to that of adjoint-valued forms. Consider a triangulation of the surfaces with $q$-dimensional cells (0-cells are vertices, 1-cells edges, etc) and the boundary operator $\partial$ mapping $q$-cells to $q-1$-cells. Instead of adjoint-value forms, we associate to each cell a vector space consisting of the adjoint representation of the group (more precisely, this can be thought of as a covariantly constant section of the associated bundle $E$ to the flat $G$-bundle). The boundary operator also acts by restricting the vector associated to a $q$-cell to its value at each boundary component made up of $q-1$-cells.

\medskip

The formal definition of the Reidemeister or combinatorial torsion is as a ratio of determinants of these boundary operators 
\beq
\tau = \frac{\sqrt{{\rm det}'\,\partial_2^\dagger \partial_2 }}{\sqrt{{\rm det}'\,\partial_1 \partial_1^\dagger}}.
\eeq
This object is independent of triangulation and reduces to the analytic torsion in the continuum limit. One can obtain a more useful version of this formula. For simplicity assume $H_2(Y,E)=0$ (these are the homology groups associated to $\partial$). The torsion is the ratio of measures 
\beq
\tau = \frac{ \alpha_2(s_1,\ldots, s_{n_2})\alpha_0(\partial t_1, \ldots, \partial t_{n_1-n_2-r},v_1,\ldots, v_k)}{\alpha_1(\partial s_1,\ldots,\partial s_{n_2}, u_1,\ldots,u_r,t_1,\ldots, t_{n_1-n_2-r})}.
\eeq
Let us define the objects involved in this formula: 
\begin{itemize}
    \item $\alpha_q(v_1,\ldots,v_{n_q})$ represents the measure of integration over the vector space living on the $q$-cells, when there are $n_q$ of them. Since these are copies of the adjoint representation, a measure on the Lie group will naturally induce such a measure. 
    \item $s_1,\ldots, s_{n_2}$ is a basis of the vector spaces of $2$-cells. We assume $H_2(Y,E)=0$ so the set of $\partial s_j$ are linearly independent.
    \item $u_1,\ldots, u_r$ where $r ={\rm dim}\, H_1 (Y,E)$ is a basis of $H_1(Y,E)$. $H_1(Y,E)$ is the cotangent bundle to the moduli space of flat bundles at the point $E$, while $H^1(Y,E)$ is the tangent bundle to moduli space. Therefore $r$ is the dimension of the moduli space of flat connections. The set $\{ \partial s_1,\ldots \partial s_{n_2}, u_1,\ldots, u_r\}$ forms a basis of $1$-cells vector spaces annihilated by $ \partial$.
    \item $t_1,\ldots, t_{n_1-n_2-r}$ are elements that complete a basis of vector spaces of $1$-cells.
    \item $v_1,\ldots, v_k$ are extra basis vectors of the vector space of $0$-cells in case $H_0(Y,E)\neq 0$.
\end{itemize}
The torsion is independent of the choice of basis $\{s_j\}$ and $\{t_j\}$. It depends on the choice of $\{u_j\}$ and $\{v_j\}$. This is reasonable since the result is a measure on $H^1(Y,E)$, the tangent space to $\mathcal{M}$, given that the inverse of a measure (since $u_j$ appears in the denominator) can be regarded as a measure on its dual space. We will see the interpretation for $\{v_j\}$ in specific examples below as related to twist parameters. The examples will hopefully clarify the abstract definition above as well.

\subsubsection*{The measure for JT gravity}

Any surface can be decomposed into three-holed spheres. Let us begin by analyzing this simple surface. What is then the moduli space of a sphere with three geodesic holes? Since there are three boundaries, flat connections will be specified by the holonomy along them, which we can call $U$, $V$ and $W$:
\vspace{5pt}
\beq
\begin{tikzpicture}[scale=0.8, baseline={([yshift=-0.1cm]current bounding box.center)}]
\draw[thick, rotate around={-30:(-2,2)}] (-2,2) ellipse (.15 and .5);
\draw[thick, rotate around={30:(2,2)}] (2,2) ellipse (.15 and .5);
\draw[thick] (0,-1) ellipse (.5 and .15); 
 \draw[thick, bend right=30] (-1.8,2.45) to (1.8,2.45);
 \draw[thick, bend left=30] (-2.2,1.55) to (-0.5,-1);
  \draw[thick, bend right=30] (2.2,1.55) to (0.5,-1);
  \node at (-2.6,2.2) {\small $U$};
   \node at (2.6,2.2) {\small $V$};
   \node at (0,-1.6) {\small $W$};
       \node at (0,1) {\small $Y$};
    \end{tikzpicture}
\eeq
Not all these matrices are independent. As the figure makes it clear, a cycle made of the three boundaries simultaneously would be contractible. This implies that the holonomies should satisfy the constraint
\beq
U V W = 1.
\eeq
Finally, we should also mod out by an overall conjugation by a group element
\beq
(U,V,W) \cong  (R UR^{-1},RVR^{-1},RWR^{-1}). 
\eeq
The constraint is evidently consistent with this identification. The three matrices give a total of $9$ parameters. The constraint $UVW=1$ provides $3$ conditions, and moding out by an overall conjugation removes $3$ parameters. This leaves a total of $3$ parameters that describe the moduli space of the three-holed sphere. These three parameters can be identified with the three geodesic lengths of the boundaries encoded in the conjugacy classes of $U$, $V$ and $W$. This is a familiar fact in hyperbolic geometry; specifying the three lengths determines a unique hyperbolic metric.

\medskip

To compute the combinatorial torsion we follow \cite{Witten:1991we,Stanford:2019vob} and pick the simplest triangulation of the surface which is
\vspace{5pt}
\beq
\begin{tikzpicture}[scale=0.8, baseline={([yshift=-0mm]current bounding box.center)}]
\draw[thick,fill=gray!60] (0,0) ellipse (2.5 and 1.5); 
 \draw[thick,fill=white] (0,-1.5) to [out=150, in=-90] (-1.3,-1.5+1.5) to [out=90, in=90] (0,-1.5) ;
  \draw[thick,fill=white] (0,-1.5) to [out=30, in=-90] (1.3,-1.5+1.5) to [out=90, in=90] (0,-1.5) ;
  \draw[fill,black] (0,-1.5) circle (0.1);
  \draw[thick,->] (1.85,-1.01) to [bend right=5] (1.85+0.1,-1.01+0.09)  ;
  \draw[thick,->] (-1,-.8) to [bend right = 30] (-1,-.8+0.01)  ;
  \draw[thick,->] (1,-.8) to [bend left= 20] (1,-.8-0.01)  ;
  \node at (2.5,-1) {\small $W$};
  \node at (-1.7,0) {\small $U$};
  \node at (1.7,0) {\small $V$};
  \node at (0,-2) {\small $P$};
      \node at (0,0.5) {\small $Y$};
    \end{tikzpicture}
\eeq
This has a single 2-cell ($Y$), three 1-cells ($U$, $V$ and $W$) and one 0-cell ($P$). Let us discuss the situation with an arbitrary gauge group $G$.  Assume that we divide the space of flat connections in $Y$ only by gauge transformations that are trivial in $P$. Then the moduli space becomes simply $G\times G$, parameterized by, say, $U$ and $V$. The definition of torsion requires the choice of the left- and right-invariant measure ${\rm vol}_G$ on the $G$ manifold. The formal definition of the combinatorial torsion gives $\widehat{\tau} = {\rm vol}(U) \cdot {\rm vol}(V)$, and the relation to $\tau$, the measure over moduli space mod all gauge transformations, is
\beq
\tau = \frac{{\rm vol}(U) \cdot {\rm vol}(V)}{{\rm vol}(R)}.
\eeq
Let us outline how to get $\widehat{\tau}$ from our definition of the torsion. We denote $\{s_1,s_2,s_3\}$ a basis for the vector space in the 2-cell (we have 3 generators since it is the dimension of $\mathrm{SL}(2,\mathbb{R})$). A basis for 1-cell vector spaces has a total of $9$ elements involving three for each boundary. Three are $\partial s_j$ and for the remaining six we choose a basis for vectors in boundary $U$, which we denote $\{u_1,u_2,u_3\}$ and boundary $V$, which we denote by $\{ v_1,v_2,v_3\}$, such that
$$\widehat{\tau}=\frac{\alpha_2(s_1,s_2,s_3)}{\alpha_1(\partial s_1, \partial s_2, \partial s_3, u_1,u_2,u_3,v_1,v_2,v_3)},$$
For a vector $s$ living on the 2-cell the boundary operator gives $\partial s = s|_U \oplus s|_V \oplus s|_W$, which implies that the contribution from $s_1,s_2,s_3$ cancels with $\partial s_1, \partial s_2, \partial s_3$. The torsion is then $$\widehat{\tau} = \alpha_1^{-1}(u_1,u_2,u_3,v_1,v_2,v_3) = {\rm vol}(U) \cdot {\rm vol}(V).$$ The remainder is just the natural measure on $U$ and $V$, divided by the measure on conjugations to obtain $\tau$. This result seems to not be symmetric under permuting the boundaries, given our choice of basis. We will see below that the final answer is always symmetric, even if in the intermediate steps are not.

\medskip 

At this point the calculation becomes straightforward; one needs to find a convenient parametrization of the moduli space and compute these group measures. We can use the freedom under conjugation to write $U=R U_0 R^{-1}$ and $V=R V_0 R^{-1}$ with 
\beq
U_0 = \updelta_1 \left(\hspace{-1mm}\begin{array}{cc}
 e^{b_1/2} & \kappa \\
0& e^{-b_1/2}
\end{array}\hspace{-1mm}\right) ,~~~V_0=\updelta_2 \left(\hspace{-1mm}\begin{array}{cc}
 e^{b_2/2} & 0 \\
1 & e^{-b_2/2}
\end{array}\hspace{-1mm}\right).
\eeq
($\updelta=\pm 1$ which we can discard now but will be important later when we include spin structures.) This choice depends on three parameters: $b_1$ and $b_2$ which can obviously be interpreted as geodesics lengths, and $\kappa$. The latter parameter should be related to $b_3$, the geodesic length of the boundary with holonomy $W$. To determine the relation write $W = R W_0 R^{-1}$ and 
\beq
W_0 = V_0^{-1} U_0^{-1}   = \updelta_1 \updelta_2 \left(\begin{array}{cc}
 e^{-(b_1-b_2)/2} & -\kappa e^{b_2/2} \\
-e^{-b_1/2} & e^{(b_1-b_2)/2}+\kappa
\end{array}\right),
\eeq
Compare the trace of $W$ with that of a diagonal matrix,
\beq
\text{Tr} W_0 = \updelta_1\updelta_2 \Big( \kappa + 2 \cosh \frac{b_1-b_2}{2} \Big) = \updelta_3\, 2 \cosh \frac{b_3}{2},
\eeq
combined with the fact that a spin structure is consistent if $\updelta_1\updelta_2\updelta_3=-1$.  We can forget about the sign here but it will be useful to keep in mind for the generalizations. This leads to
\beq
\kappa = - 2 \cosh \frac{b_3}{2} - 2 \cosh \frac{b_1-b_2}{2}.
\eeq
 Finally we need to make a choice of a group measure. We can represent an element in the algebra $\mathfrak{sl}(2,\mathbb{R})$ by a $2\times 2$ matrix expanded in a basis $\{e,f,h\}$ as $x=\Big(\begin{array}{cc}
 x_h & x_e \\
x_f & -x_h
\end{array}\Big) = x_e \,e + x_f \, f + x_h \, h$. The measure derived from the inner product $| x|^2 = 2 \text{Tr} \, x^2 $ is just $4 \d x_e \d x_f \d x_h$. On a group element $U$ we can write this measure as 
\beq
\text{vol} (U) = 4 (U^{-1} \d U)_e(U^{-1} \d U)_f(U^{-1} \d U)_h.
\eeq
Now we have all the ingredients we need to evaluate the torsion of the three-holed sphere which we leave as an exercise, or you can read section 3.4.2 of \cite{Stanford:2019vob}. With this parametrization of $U$ and $V$, and the definition of the measure as ${\rm vol}(U) \cdot {\rm vol}(V)/{\rm vol}(R)$, show that 
  \beq
\tau = 8 \sinh \frac{b_1}{2} \sinh \frac{b_2}{2} \, \d b_1 \,\d b_2 \, \d  \kappa
  \eeq
Using the relation between $\kappa$ and $b_3$ we can rewrite the torsion as:
\beq
\tau = 8 \sinh \frac{b_1}{2} \sinh \frac{b_2}{2} \sinh \frac{b_3}{2}\,  \d b_1 \, \d b_2 \, \d b_3
\eeq
The final answer for the torsion turns up to be symmetric on the three boundaries as anticipated. The prefactor looks funny. We expect the measure over the three-holed sphere to be independent of the boundary geodesic lengths. To address this we need to consider how the pair-of-pants are supposed to be glued to each other. 

\subsubsection*{Gluing and torsion of a circle}
What happens when we glue together two manifolds $Y_1$ and $Y_2$ with a common boundary circle $S_{12}$? In quantum field theory, gluing is implemented by multiplying the path integrals over $Y_1$ and $Y_2$ and summing over physical states propagating along the common boundary $S_{12}$. In the approach via the torsion to $BF$ theory, the appropriate gluing procedure is to multiply the torsions of $Y_1$ and $Y_2$ and divide by the torsion of the circle $S_{12}$:
\beq
\frac{\tau_{Y_1} \cdot \tau_{Y_2}}{\tau_{S_{12}} }.
\eeq
A detailed explanation of the procedure can be found in section 4 of \cite{Witten:1991we}. Since the torsion is a measure that includes fluctuations of the boundary holonomies, the rough idea is that we need to divide by the circle torsion to avoid overcounting. Otherwise gluing two pair-of-pants would lead to unwanted terms involving $(\d b)^2$. 

\medskip

Let us now discuss the torsion of the circle. Consider a flat connection on a circle with holonomy $U=\text{diag}(e^{b/2},e^{-b/2})$. This has an obvious `triangulation', a base point in the circle, and the circle itself. The general procedure is slightly subtle and is outlined in section 3.4.3 of \cite{Stanford:2019vob}. For an element of the algebra $t$ on the 1-cell, the boundary operator $\partial$ acts as $\partial t = U t U^{-1} - t$, due to the holonomy around the circle. The definition of the torsion gives 
\beq
\tau = \frac{\alpha_0(\partial t_1, \partial t_2, v) }{ \alpha_1(\underbrace{t_1,t_2}_{\text{off-diagonal matrices}},\underbrace{u}_{\text{diagonal matrix}})} = | {\rm det}'\, \partial |\, \frac{\alpha_0(v)}{\alpha_1(u)}.
\eeq
We parameterized the vectors in the 1-cell by $t_1,t_2$, the two off-diagonal $2\times 2$ matrices and $u$, a diagonal matrix that is therefore a zero-mode of $\partial$. $v$ is a matrix that commutes with $U$ and its needed to complete the base of vectors on the 0-cell. The determinant over non-zero-modes, after a simple calculation (The eigenvalues of $\partial$ are $e^{\pm b}-1$ and $0$), leads to
\beq
| {\rm det}'\, \partial |= 4 \sinh^2 \frac{b}{2}.
\eeq
What about $\alpha_0(v)/\alpha_1(u)$? The denominator arises from the tangent space to the moduli space of flat connections and should be associated to changes in the geodesic length through $U={\rm diag}(e^{b/2},e^{-b/2})$. The numerator arises from matrices in $\SL(2,\mathbb{R})$ that commute with $U$, which naturally appears when gluing through parallel transport across the circle and we called it ${\rm diag}(e^{\varrho/2},e^{-\varrho/2})$. The ratio of measures is then naturally $ \d b\, (\d \varrho)^{-1}$. The final answer is
\beq
\tau_S = 4 \sinh^2\Big( \frac{b}{2}\Big)\, \, \d b\, \cdot\, (\d \varrho)^{-1}
\eeq
This is great, the factor of $\d b$ will cancel the extra unwanted term in the product of two three-holed-sphere torsion while the twist parameter $\d \varrho$ will replace it.

\subsection{JT gravity as a matrix integral}

Having determined the building blocks of the measure over hyperbolic surfaces relevant for JT gravity we can complete the calculation and show how the result can be reproduced by a matrix integral. Let us begin with surfaces without boundaries. Any closed oriented surface $\Sigma$ of genus $g$ can be assembled by gluing together a set $T$ of $2g-2$ three-holed spheres $Y_t$, $t\in T$. These three-holed spheres have to be glued along a set $C$ of $3g-3$ circles $S_c$, $c\in C$. Two three-holed spheres (or two boundaries of the same three-holed sphere) are glued along each $S_c$. Then
\beq
\mu_{g} = \prod_{t\in T} \tau_{Y_t} \,\, \prod_{c\in C} \frac{1}{\tau_{S_c}}.
\eeq
The result is quite simple. Each circle of length $\d b$ bounds two three-holed sphere whose torsion produces a factor $(\d b)^2$. The torsion of that circle replaces one $\d b$ by $\d \varrho$, leading to a factor of $\d b \, \d \varrho$ for each circle. Moreover one can easily see that all factors of $2 \sinh b/2$ nicely cancel between the three-holed spheres and the circles. The final torsion of a closed surface of genus $g$ and no boundaries $n=0$ is given by
\beq
\mu_g= \prod_{i=1}^{3g-3} \d b_i \, \d \varrho_i 
\eeq
This is precisely the Weil-Petersson volume form covered the first week! We could have derived this from a symplectic approach but, as we will see, the torsion is more powerful. The JT gravity partition function on closed hyperbolic surfaces is then given by
\beq\label{eq:JTwobdy}
Z_{g,0} = \int_{\mathcal{M}_{g,0}} \prod_{i=1}^{3g-3} \d b_i \, \d \varrho_i = V_{g,0},
\eeq
where $V_{g,n=0}$ is the corresponds Weil-Petersson volume. The integral is done over $\mathcal{M}_{g,0}$, the moduli space of hyperbolic surfaces of genus $g$. 

As we mentioned in the previous section, the connection between $BF$ theory and gravity is subtle. First we need to restrict to the right component of flat connection, the Teichmuller space $\mathcal{T}_{g,0}$. We achieved this by assuming that all holonomies are in the hyperbolic conjugacy class. Second we need to mod out by the mapping class group (the measure we derived on $\mathcal{T}$ naturally descends to a measure on $\mathcal{M}$). Two surfaces with very different values of $b_i$ and $\varrho_i$ might be non-trivial equivalent under a large diffeomorphism. Fortunately, once we get to this point, we can use the results from Mirzakhani to evaluate $V_{g,n}$ using the recursion relation  \cite{Mirzakhani:2006eta,mirzakhani2007simple} that is covered in Giacchetto and Lewanski's lectures \cite{GiacchettoLesHouchesNotes}.

\medskip

The result \eqref{eq:JTwobdy} implies that JT gravity partition function without boundaries is equal to the Weil-Petersson volumes, which combined with the result of Eynard and Orantin \cite{Eynard:2007fi} implies that they are computed by a double-scaled matrix integral! Precisely the spectral curve derive from the JT gravity disk partition function is the one identified by Eynard and Orantin as required to reproduce the Weil-Petersson volumes. Before going into more details let us introduce boundaries.

\medskip

We would like to work in the presence of NAdS boundaries introduced in the previous lecture. This cannot be done in the language of the torsion or even $BF$ theory in a rigorous way, but can be inferred from previous results. For example, while $Z_{\rm JT}^{\rm trumpet}(\beta,b)$ is a number, the path integral with torsion boundary conditions $\widetilde{Z}_{\rm JT}^{\rm trumpet}(\beta,b)\, \d b$ should be a measure of integration over $b$. To determine it we can compare the two-boundary wormhole
\beq
Z_{\rm JT}^{\rm trumpet}(\beta_1,b)\,\, \d b \d \varrho \, \,Z_{\rm JT}^{\rm trumpet}(\beta_2,b),
\eeq
with the quantity with torsion boundary conditions
\beq
\widetilde{Z}_{\rm JT}^{\rm trumpet}(\beta_1,b)\, \d b \, \frac{1}{\tau_S} \, \widetilde{Z}_{\rm JT}^{\rm trumpet}(\beta_2,b)\, \d b.
\eeq
Comparing both quantities we infer that the trumpet path integral with torsion boundary conditions is given by
\beq
\widetilde{Z}_{\rm JT}^{\rm trumpet}(b) = Z_{\rm JT}^{\rm trumpet} \cdot 2 \sinh \frac{b}{2}.
\eeq
We are ready to write the final answer for the JT gravity path integral with $\text{NAdS}$ boundaries. First of all, the argument for the case without boundaries still implies that the integration over all internal geodesics is done with a measure $\prod_i \d b_i\, \d \varrho_i$. What about geodesic connected to trumpets? The torsion of the three-holed sphere, together with the trumpet and circle torsions, will include  term
\beq
\prod_{i\in C_{\rm int}}\d b_i\, \d \varrho_i \cdot \prod_{e \in C_{\rm ext}} \underbrace{2 \sinh \frac{b_e}{2} \d b_e }_{\text{Leftover from torsion of three-holed sphere}}\,\, \frac{1}{\tau_{S_e}} \,\, \underbrace{Z_{\rm JT}^{\rm trumpet} \cdot 2 \sinh \frac{b_e}{2} \d b_e}_{\text{Trumpet with torsion bdy}}
\eeq
where $C_{\rm int}$ is the set of internal geodesics while $C_{\rm ext}$ is the set of geodesics connected to external boundaries. The JT gravity partition function is the integral of the moduli space of hyperbolic surfaces using this measure. To relate it to the quantity computed by Mirzakhani, we can write the final answer as
\beq
Z_{g,n}(\beta_1,\ldots,\beta_n) =  \prod_{e\in C_{\rm ext}} \int_0^\infty b_e\, \d b_e Z_{\rm JT}^{\rm trumpet} (\beta_e,b_e) \,\, \underbrace{ \int_{\mathcal{M}_{g,\vec{b}}} \, \prod_{i\in C_{\rm int}}\d b_i\, \d \varrho_i}_{\text{Weil-Petersson volume $V_{g,n}(b_1,\ldots,b_n)$}}.
\eeq
We denote by $\mathcal{M}_{g,\vec{b}}$ the moduli space with geodesic boundaries to distinguish from $\mathcal{M}_{g,n}$, the moduli space with $n$ punctures. The internal Weil-Petersson volume does not care about the twist parameter used for the external gluing and is therefore independent of it. This allows us to perform the integral over $\varrho_e$ rather trivially and obtain factors of $b_e$.

\medskip

The answer in its final form becomes, after writing explicitly the trumpet path integral,
\beq\label{eq:ZgnintermsofV}
Z_{g,n}(\beta_1,\ldots,\beta_n) =  \left(\prod_{e=1}^n \int_0^\infty b_e \d b_e\,\, \frac{e^{-b^2/(4\beta)}}{\sqrt{4 \pi \beta}}\right) \,\, V_{g,n}(b_1,\ldots,b_n). 
\eeq
Now all the ingredients are explicit and known. The Eynard-Orantin result now implies that 
\beq
Z_{g,n} = \left\langle {\rm Tr} \, e^{-\beta_1 H} \ldots {\rm Tr} \, e^{-\beta_n H} \right\rangle^{\rm conn.}_{g,n}.
\eeq
The GPI of JT gravity on NAdS boundaries is equal, order by order in the topological expansion, to the product of thermal partition functions averaged over the Hamiltonian! The result boils down to the theorem of Eynard and Orantin, but the crucial new ingredient is to show precisely how the volumes that Mirzakhani computed arise from the GPI.

\subsubsection*{Outline of the proof}

For completeness we will outline some intermediate steps in the proof. Let us begin from the matrix model side. The loop equations are more naturally written in terms of resolvents
\beq\label{eq:ZvsR}
R(x) = \text{Tr} \, \frac{1}{x-H},~~~\Rightarrow~~~R(x) = - \int_0^\infty \d \beta \, e^{\beta x} \, Z(\beta).
\eeq
The ensemble average of a product of $n$ such resolvents has a topological expansion in the 't Hooft limit as well as in the double-scaling limit. We denote the genus $g$ contribution to the connected correlator by $R_{g,n}(x_1,\ldots, x_n)$. For the GUE ensemble the loop equations, valid for any spectral curve or equivalently any matrix potential, are 
\beq\label{sdsds}
2xy(x) R_{g,n+1}(x,I) + x F_{g,n}(x,I) = \text{(analytic in $x$)},
\eeq
where $I=\{ x_1,\ldots, x_n\}$ and
\bea
F_{g,n}(x,I) &=& R_{g-1,n+2}(x,x,I) + \sum_{\rm stable} R_{h_1,|I_1|+1}(x,I_1) R_{h_2,|I_2|+1}(x,I_2)\nonumber\\
&&+ 2 \sum_{k=1}^n \Big( R_{0,2}(x,x_k) + \frac{1}{2} \frac{1}{(x-x_k)^2}\Big) R_{g,n}(x,I/x_k).
\ea
The sum in the second term on the first line involves subsets $I_1 \cup I_2 = I$ and genus $h_1+h_2=g$ that are stable, meaning that $R_{0,1}$ or $R_{0,2}$ never appears. A dispersion relation argument that effectively discards the analytic part of \eqref{sdsds} gives
\beq
R_{g,n+1}(x,I) = \int_{\mathcal{C}} \frac{\d x'}{2\pi \i} \frac{1}{x'-x} \frac{\sqrt{\sigma(x')}}{\sqrt{\sigma(x)}} \frac{F_{g,n}(x',I)}{2 y(x')}.
\eeq
For a leading density of states with support between $a_-$ and $a_+$ we define $\sigma(x) = (x-a_-)(x-a_+)$. For JT gravity we have $a_-=0$ and in the double scaling limit $a_+\to\infty$. The contour $\mathcal{C}$ surrounds the cut. In the case of the GUE ensemble one can show that the integral reduces to a pole evaluated at the endpoints $x=a_\pm$, which for the JT gravity spectral curve would reduce only to the edge at $x=0$. This equation determines recursively order-by-order all terms in the topological expansion.

Lets consider a double-scaled spectrum with $a_-=0$ for simplicity. We can define the following quantity contructed out of the matrix integral resolvent
\beq\label{VRMTvsR}
V_{g,n}^{\rm RMT}(b_1,\ldots,b_n) = \int_{\varepsilon + \i \mathbb{R}} R_{g,n}(-z_1^2, \ldots, -z_n^2) \prod_{j=1}^n \frac{\d z_j}{2\pi \i}\, \frac{-2 z_j}{b_j} \, e^{b_j z_j},
\eeq
the contour being the one corresponding to an inverse Laplace transform. The variable $z$ is defined such that $x=-z^2$. The quantity $V_{g,n}^{\rm RMT}$ satisfies the following dispersion relation, derived from an inverse Laplace transform of the loop equations
\bea
b V_{g,n+1}^{\rm RMT}(b,B) &=& \frac{1}{2} \int_0^\infty b'\d b' \int_0^\infty b'' \d b'' \, D(b'+b'',b) \nonumber\\
&&\times \left( V_{g-1,n+2}^{\rm RMT}(b',b'',B) + \sum_{\rm stable} V_{h_1, |B_1|+1}^{\rm RMT}(b',B_1) V_{h_2,|B_2|+1}^{\rm RMT}(b'',B_2) \right)\nonumber\\
&&+ \frac{1}{2} \sum_{k=1}^{|B|} \int_0^\infty b'\d b'\, \Big( D(b'+b_k,b) + D(b'-b_k,b) \Big) V_{g,n}^{\rm RMT}(b',B/b_k),\label{eqn:MTRMTM}
\ea
where $B=\{b_1,\ldots, b_n\}$ and the sum in the second line is over subsets $B_1 \cup B_2 = B$ and genus $h_1+h_2=g$. The kernel appearing in the integrals is related to the spectral curve by 
\beq
D(x,y) = - \int_{\varepsilon+\i \mathbb{R}} \frac{\d z}{2\pi \i} \, \frac{e^{- x z} \sinh\, yz}{z y(-z^2)}.
\eeq
A derivation for an arbitrary spectral curve can be found in Appendix B of \cite{Turiaci:2023jfa}. Comparing \eqref{eqn:MTRMTM} with the results covered in week one, more precisely section 4 of Giacchetto and Lewanski's chapter \cite{GiacchettoLesHouchesNotes} and also Bouchard's \cite{bouchard2024leshoucheslecturenotes}, this recursion relation takes precisely the same form as Mirzakhani's one for the Weil-Petersson volumes. Matching the precise form of the kernels requires precisely the spectral curve of JT gravity
\beq
D(x,y) = - 4\pi \int_{\varepsilon+\i \mathbb{R}} \frac{\d z}{2\pi \i} \, \frac{e^{- x z} \sinh\, yz}{z \sin 2\pi z}= y - \log \left(\frac{\cosh^2 \frac{x+y}{4}}{\cosh^2 \frac{x-y}{4}}\right).
\eeq
This was the insight of \cite{Eynard:2007fi} and shows that $V_{g,n}^{\rm RMT}$ constructed out of the matrix model resolvent is equal to $V_{g,n}$, the Weil-Petersson volume of moduli spaces of hyperbolic surfaces, when the spectral curve is taken to be the one of JT gravity.

The final step of the proof is to show that the integral transform \eqref{VRMTvsR} between $V^{\rm RMT}_{g,n}$ and $R_{g,n}$ is the same one we would deduce from gravity. Since the gravity answer is given in terms of partition functions $Z_{g,n}$, we can translate to resolvents by applying the integral transform \eqref{eq:ZvsR} to \eqref{eq:ZgnintermsofV}. The result is
\beq
R_{g,n}(-z_1^2,\ldots,-z_n^2) = (-1)^n \left(\prod_{e=1}^n \int_0^\infty  \d b_e\,\, \frac{e^{-b_e z_e}}{2 z_e}\right) \,\, V_{g,n}(b_1,\ldots,b_n)
\eeq
This relation is precisely the same as \eqref{VRMTvsR}, completing the proof. This implies that the GPI of JT gravity is equal to a matrix integral, order by order in perturbation theory. This is the achievement of Saad, Shenker and Stanford in \cite{Saad:2019lba}.

\section{Generalizations of the JT/RMT duality}\label{sec:generaliz}

\subsection{Dilaton-gravity as a matrix integral}

Matrix integrals on a given symmetry class are parameterized by one function, the matrix potential, or equivalently, the spectral curve. 2d dilaton-gravity theories are also parameterized by a function, the dilaton potential. Could the two theories be related beyond the specific example of JT gravity? If so, what is the relation between the spectral curve and the dilaton potential? 

\medskip

Under some assumptions on the dilaton potential, both questions were answered positively with Maxfield \cite{Maxfield:2020ale} and independently by Witten \cite{Witten:2020wvy} for a specific class of theories. We will outline these results here and mention some  recent progress. 

\medskip 

Let us recall the specific form of the dilaton-gravity action in terms of the dilaton potential
\beq
    I = - \frac{S_0}{4\pi} \int_{M} \sqrt{g} R- \frac{1}{2} \int_{M} \sqrt{g}(\Phi R + U(\Phi)) + I_{\rm bdy}.
\eeq
The boundary terms are not affected by the precise dilaton potential. The work of \cite{Maxfield:2020ale,Witten:2020wvy} applies to dilaton potentials of the following form
\beq
U(\Phi) = 2 \Phi + \sum_I \, \lambda_I \,\,e^{- (2\pi - \alpha_I) \Phi},~~~~0\leq \alpha_I \leq \pi.
\eeq
What is special about this choice? Consider the theory with a single exponential for simplicity. Since we already solved the theory with $\lambda=0$ it is reasonable to explore how a Taylor expansion in $\lambda$ would look like. To each order in $\lambda$ the result would look like the JT gravity partition function with an insertion of 
\beq
\frac{\lambda^k}{k!}\,\, \sqrt{g} \, \int [\d g\, \d \Phi ]\,\, \Big(\prod_{j=1}^k\int_M \d^2 x_j\sqrt{g} \,e^{- (2\pi -\alpha)  \Phi(x_j)} \Big) \, e^{-I_{JT}} \,.
\eeq
Recall that the JT gravity action is linear in $\Phi$ which acts as a Lagrange multiplier setting $R=-2$. Assume we can commute the two integrals, performing first the JT gravity path integral with an insertion of $\prod_j \, e^{-(2\pi - \alpha)\Phi(x_j)}$ with fixed positions. The effect of the exponential insertions become very simple, as observed earlier in \cite{Mertens:2019tcm}, replacing the constant negative curvature condition by
\beq
R+2 = 2\sum_{j=1}^k (2\pi -\alpha) \, \delta(x_j).
\eeq
This implies that the GPI at order $k$ localizes into hyperbolic geometries with constant curvature and with $k$ conical singularities, with opening angle $\alpha$, located at $x_1,\ldots,x_k$. For reference, a cusp corresponds to $\alpha=0$ while the conical deficit disappears when $\alpha=2\pi$. Notice that in the end we are required to integrate over the positions $x_j$ which become part of the moduli space of hyperbolic surfaces with conical defects. The prefactor of $1/k!$ just guarantees that defects are treated as indistinguishable. The overall factor of $\lambda^k$ means that each defect insertion comes with a weight $\lambda$. 

\medskip

To solve the theory we should first determine the JT gravity path integral in the presence of a fixed number $k$ of defects. Next we should sum over all possible values of $k$. It is surprising that the second step can be performed in closed form for the disk allowing us to extract the deformed spectral curve. Moreover, we will also show the result is still dual to a matrix integral with that density of states.

\medskip

We can write the GPI for dilaton gravity, over connected spacetimes, as a topological expansion
\beq
Z_{\rm grav}(\beta_1,\ldots,\beta_n) = \sum_{g=0}^\infty e^{S_0 \chi} Z_{g,n}(\beta_1,\ldots,\beta_n)
\eeq
with each term being given in turn by a sum over defects
\beq
Z_{g,n}(\beta_1,\ldots,\beta_n) = \sum_{k=0}^\infty \frac{\lambda^k}{k!}\, Z_{g,n,k} (\beta_1,\ldots, \beta_n; \underbrace{\alpha,\ldots,\alpha}_{\text{k terms}}).
\eeq
$Z_{g,n,k}$ is the path integral on surfaces with genus $g$ and $k$ defects. In this case we consider a single defect species so all $\alpha$'s are the same but this can be generalized in a straightforward fashion. This double expansion in genus and number of defects can be represented in the following figure, taken from \cite{Maxfield:2020ale},
$$
\includegraphics[scale=0.6]{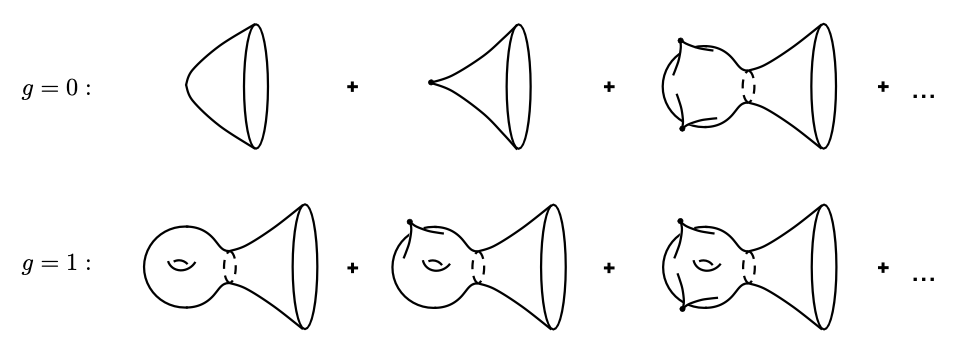}
$$
We need to evaluate $Z_{g,n,k}$. Let us begin with the simplest case of the disk with a defect $Z_{0,1,1}(\beta; \alpha)$. In this case the boundary graviton theory reduces to the Schwarzian mode in the space 
\beq
\text{Diff}(S^1)/\U(1),
\eeq
since the defect breaks the isometries of the empty hyperbolic disk. We can apply the localization theorem to fixed point again and evaluate the quantum effects from the rotation angles. We obtain \cite{Mertens:2019tcm}
\beq
Z_{0,1,1}(\beta) = \frac{1}{\sqrt{4\pi\beta}} e^{\frac{ \alpha^2}{4\beta}}.
\eeq
Notice that this is same as the trumpet partition function under the replacement
\beq
b \to \i \alpha.
\eeq
This is not a coincidence. We can compare the geodesic hole to the defect using the $BF$ formulation. The holonomy of a hole of length $b$ and a conical singularity of opening $\alpha$ are
\vspace{5pt}
\bea
U_b =\updelta \exp{\left(\begin{array}{cc}
     b/2 &0 \\
     0 & -b/2 
\end{array}\right)} ,~~~~U_\alpha =\updelta \exp{\left(\begin{array}{cc}
     0 & \alpha/2 \\
     - \alpha/2 & 0 
\end{array}\right)}
\ea
These two holonomies are conjugate to each other in $\SL(2,\mathbb{R})$ if we identify $b= \i \alpha$. This is the simplest fact indicating that one might be able to obtain results for defects by analytic continuation on geodesic lengths of holes.

\medskip

The relation between holes and defect continue to hold in more complicated surfaces if and only if all opening angles satisfy $0\leq \alpha \leq \pi$. In fact, it was proven by Tan Wong and Zhang \cite{tan2004generalizationsmcshanesidentityhyperbolic}, and further developed by Do and Norbury \cite{do2006weilpeterssonvolumesconesurfaces}, that the Weil-Petersson volumes of moduli spaces of hyperbolic surfaces with $n$ holes of length $\vec{b}$ and $k$ defects $\vec{\alpha}$ in this range are given by
\beq
V_{g,n,k}(\vec{b}; \vec{\alpha}) = V_{g,n+k}(b_1,\ldots, b_n,  \i \alpha_1, \ldots, \i \alpha_k).
\eeq
The quantity in the right hand side is the corresponding Weil-Petersson volume with $n+k$ geodesic boundaries. Besides this, it is important to assume $0\leq \alpha \leq \pi$ when gluing the surface to the trumpets that connect with NAdS boundaries since it guarantees that all NAdS boundaries are homologous to geodesics without encountering any defect. This directly implies that the partition function can be obtained by gluing 
\beq
Z_{g,n,k}(\beta_1,\ldots, \beta_n) =\left( \prod_{e=1}^n \int_0^\infty b_e\,\d b_e\, Z_{\rm JT}^{\rm trumpet}(\beta_e,b_e) \right)  V_{g,n,k}(\vec{b}; \vec{\alpha})
\eeq
\vspace{5pt}
The volumes are computed naturally for distinguishable defects.

\subsubsection{The deformed spectral curve}

Having solved the theory at all orders in topology and in $\lambda$ we turn to the following question; what is the deformed leading order density of states? This arises from surfaces with genus $g=0$ but an arbitrary number of defects. This would determine the deformed spectral curve of the dual matrix model, or equivalently the matrix potential, necessary to implement the loop equations. Therefore this calculation is necessary to even propose a candidate matrix model associated to a given dilaton-potential. 

\medskip

Let us consider the following question first. Does a theory gravity make sense if we restrict the number of defects? For example, consider a theory of gravity with either none or one defect. Can the theory be holographic, i.e. still be dual to a random matrix model? The correction to the density of states from one defect gives
\beq
e^{-S_0}\rho_{g=0}(E) = \frac{\sinh 2\pi \sqrt{E}}{4\pi^2} + \lambda \frac{\cosh (\alpha \sqrt{E})}{2\pi \sqrt{E}} \sim_{E\to 0} \frac{\lambda}{2\pi \sqrt{E}}.
\eeq
This behavior is inconsistent with a random matrix in the GUE ensemble! Only powers of $E^{n/2}$ with $n\geq 1$ are allowed for such an ensemble near threshold. Therefore JT gravity with none or one defect is a reasonable-looking theory of gravity which is not holographic (in the sense of being dual to at least a matrix integral). This simple observation had an important consequences for investigations of pure 3d gravity as observed in \cite{Maxfield:2020ale}. The Maloney-Witten partition function, when interpreted as being on 2d surface fibered over a circle, is precisely the analog of including a single defects in JT gravity. 

\medskip

Does the sum over arbitrary numbers of such defects solve the problem? The answer is that contributions with more defects diverge faster and faster near threshold, and a resummation is necessary to decide what the actual behavior at the edge is. To derive this property it is useful to recall that when one of the boundaries is large, the Weil-Petersson volume is approximated
\beq
V_{0,k+1}(b_0, \vec{b}) = \frac{1}{(k-2)!} \frac{b_0^{2k-4}}{2^{k-2}} + \ldots, ~~~~b_0\gg 1
\eeq
This implies that 
\beq
Z_{0,1,k}(\beta) = \frac{1}{\sqrt{2\pi}} (2\beta)^{k-3/2} + \ldots, 
\eeq
which after resummation leads to
\beq
Z(\beta) = e^{S_0} \sum_{k=0}^\infty \frac{\lambda^k}{k!} Z_{0,1,k}(\beta) = \frac{1}{4 \sqrt{\pi}\beta^{3/2}} e^{2 \lambda \beta} + \ldots.
\eeq
In terms of the density of states this implies
\beq
\rho(E) \sim e^{S_0}\frac{1}{2\pi} \sqrt{2(E-E_0)} ,~~~E_0 = -2\lambda + O(\lambda^2). 
\eeq
Therefore the singularity at $E=0$ is simply signaling the fact that the deformation includes a shift in the threshold energy, but the edge is still a square-root one! As proposed in \cite{Maxfield:2020ale} the analog procedure in three-dimensional gravity that would correct the Maloney-Witten partition function is to sum over Seifert manifolds.

\medskip 

So far we checked the threshold behavior, but can we determine the full spectral curve? To do this we can use the following exact formula for Weil-Petersson volumes derived in \cite{Mertens:2020hbs}\footnote{The derivation can be done in the language of the orthogonal polynomial method that C. Johnson described in his lecture \cite{JohnsonLesHouchesNotes}. The equation for $u(x)$ is the tree-level string equation. The formula for $V_{g,n}$ arises from a general formula for genus-zero correlators for any double-scaled matrix integral derived in \cite{Moore:1991ir}.}
\beq
V_{0,n}(b_1,\ldots,b_n) = \frac{(-1)^{n-1}}{2} \frac{\d^{n-3}}{\d x^{n-3}}\left[u'(x) \prod_{i=1}^n J_0\left(b_i\sqrt{u(x)}\right) \right]\Big|_{x\to0},~~~\frac{\sqrt{u}}{2\pi} I_1 ( 2\pi \sqrt{u})= x
\eeq
we can derive the exact density of states by summing over an arbitrary number of defects on the disk. The derivation is quite involved and can be found in section 3 of \cite{Maxfield:2020ale}. This expression and similar ones at higher genus were later derived by Eynard and Lewanski using independent methods \cite{eynard2023naturalbasisintersectionnumbers}. This can be used to evaluate $Z_{0,1,k}$ explicitly, and moreover the expression can be resummed using Lagrange's reversion theorem, leading to
\beq
\rho(E) = \frac{e^{S_0}}{2\pi} \int_{E_0}^E\,\frac{\d u}{\sqrt{E-u}} \, \frac{\d F(u)}{\d u},
\eeq
where
\beq
F(u)= \frac{\sqrt{u}}{2\pi} I_1(2\pi \sqrt{u}) + \sum_I \lambda_I \, I_0(\alpha_I \sqrt{u}).
\eeq
The threshold energy is the largest root of $F(E_0)=0$ and one can show that the edge is a square root, as universally expected for a random matrix spectrum. At high energies, $E\gg1$ the density of states grows exponentially as $e^{2\pi \sqrt{E}}$ which corresponds to the classical regime of JT gravity. The function $F(u)$ is precisely the tree-level string equation that appeared in C. Johnson's lectures \cite{JohnsonLesHouchesNotes}. Even though the expression for $\rho(E)$ is quite complicated, it is nice that the defect species are additive in the string equation. We are now ready to propose a random matrix dual; it is the GUE ensemble with a matrix potential producing a leading density of states reproducing this expression.

\medskip

As a side comment, it is interesting that the density of states can become negative in certain ranges of the deformation. This was noted in \cite{Maxfield:2020ale,Witten:2020wvy} and further studied by F. Rosso and C. Johnson in \cite{Johnson:2020lns}. In \cite{Rosso:2021orf} with Rosso we showed that in some simpler settings, namely $\mathcal{N}=1$ supergravity, the matrix model indicates a transition to a two-cut phase. This leaves an interesting open question: What is the gravity interpretation of such phase transitions?

\subsubsection{Proof of the duality between deformations of JT gravity and RMT}
For simplicity consider a single defect species. To prove the duality to all orders in the genus expansion, it is useful to write the partition function in a slightly different way. First we remind that the resolvent and Weil-Petersson volumes are related by
\beq
W_{g,n}(\vec{z}) = \left( \prod_{e=1}^n\,\int_0^\infty b_e\,\d b_e e^{-b_e z_e}\right) V_{g,n}(\vec{b}).
\eeq
where $W_{g,n}(\vec{z})= (-2 z_1)\ldots (-2 z_n) R_{g,n}(-z_1^2,\ldots, -z_n^2)$. We can write the deformed $W_{g,n}(\vec{z};\lambda)$ and evaluate the derivative with respect to $\lambda$ at $\lambda=0$. These are the path integrals with defects precisely. Some simple manipulations lead to
\beq
\frac{\d^k}{\d \lambda^k} W_{g,n}(\vec{z})\Big|_{\lambda=0} = \left( \int_0^\infty \prod_{e=1}^k b_e \, \d b_e e^{-b_e z_e} \right) V_{g,n+k}(b_1,\ldots,b_n ,\underbrace{ \i \alpha, \ldots, \i \alpha}_{\text{$k$ terms}})
\eeq
We can combine this with an inverse Laplace transform 
\beq
V_{g,n}(\vec{b}) = \left(\prod_{e=1}^n \,\int_{\mathcal{C}} \frac{\d z}{2\pi \i} \, \frac{e^{b_e z_e}}{b_e} \, \right) W_{g,n}(\vec{z}).
\eeq
This leads to
\beq
\frac{\d^k}{\d \lambda^k} W_{g,n}(\vec{z})\,\Big|_{\lambda=0} = \left(  \prod_{e=1}^k\, \int_{\mathcal{C}} \frac{\d \tilde{z}_e}{2\pi \i} \frac{\sin( \alpha \tilde{z}_e)}{\alpha}\, \right) W_{g,n+k}(\vec{z} , \widetilde{z}_1,\ldots,\widetilde{z}_k),
\eeq
where we write $e^{\i \alpha \tilde{z}_e}/\i\alpha$ in terms of the sine for convenience; it guarantees the formula is valid for $(g,n)=(0,1)$ as well. We can take as integration contour a curve that encircles the cut. Finally this is equivalent to the relation at finite $\lambda$
\beq
\frac{\d W_{g,n}(\vec{z};\lambda)}{\d \lambda} = \oint_{\mathcal{C}} \frac{\d \tilde{z}}{2\pi \i} \, f(\tilde{z}) \, W_{g,n+k}(\vec{z},\tilde{z};\lambda),~~~f(z)= \frac{\sin(\alpha z)}{\alpha}
\eeq
This is precisely of the form relevant for the ``deformation theorem'' of Eynard and Orantin \cite{Eynard:2007fi}, which states that deformations that take this form automatically satisfy the topological recursion and therefore the matrix integral loop equation. (The theorem is even more general and allows the integrand, which in our case is $\sin(\alpha z)/z$ to also depend on $\lambda$. We did not find an application of this generalization to gravity.)

\subsubsection{Recent progress on blunt defects and applications}

So far we only considered deformations where the defects are sharp enough with $0\leq \alpha \leq \pi$. When the defects are blunt, and fall in the range $\pi < \alpha \leq 2\pi$, the previous results are not valid. This is relevant when considering dilaton-potentials that affect the classical solution for the black hole geometry and not only the quantum-corrected regime. To achieve this one needs to take $\alpha-2\pi$ small. This has a few applications. First, we would like to know, in the classical limit, how the backreaction from the dilaton potential arises from resumming a gas of defects corresponding to singular geometries. As another application, dilaton potentials corresponding to blunt defects can have interesting singularities in Lorentzian signature \cite{Kruthoff:2024gxc}. 

\medskip

In the absence of boundaries, the problem with blunt defects is essentially to take care of situations where defects can merge such as
$$
\includegraphics[scale=0.2]{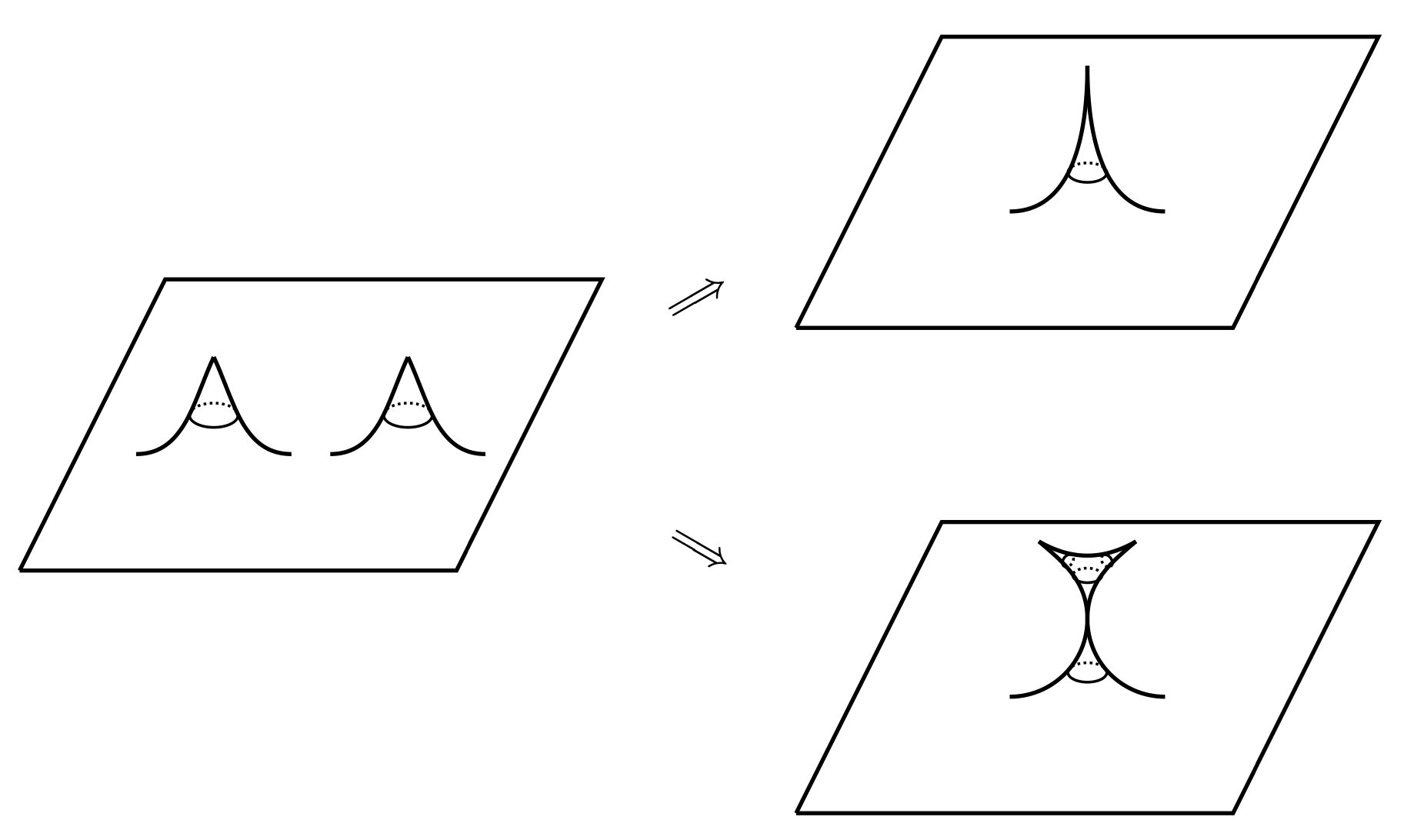}
$$
The Deligne-Mumford prescription for compactifying moduli space treats the merging of defects in the way depicted in the bottom adapted to sharp defects. Therefore, the merger of blunt defects leading to a single effective defect, without cusps, is not correctly treated by this approach. This is illustrated by the fact that the Weil-Petersson volumes do not behave in a reasonable way when considering blunt defects. For example the sphere with four defects would give
\bea
V_{0,0,4}(\vec{\alpha}) =?\, \frac{4\pi^2 - \alpha_1^2-\alpha_2^2-\alpha_3^2-\alpha_4^2}{2},
\ea
However, even when obeying the Gauss-Bonnet constraint 
\beq
\frac{1}{2}\int R + \sum_j (2\pi-\alpha_j) = 2\pi \chi,
\eeq
which implies that $\sum_j (2\pi - \alpha_j) \geq 4\pi$, one can obtain negative answer, for example for $\alpha_1=\alpha_2=0$ and $\alpha_3=\alpha_4>\sqrt{2} \pi$. Another evidence is that when we put $\alpha=2\pi$, turning the defect into nothing, we do not recover the answer with one defect less! 

\medskip

How can we correct this problem for blunt defects? In a paper with Usatyuk and Weng \cite{Turiaci:2020fjj} we proposed an answer based on a limit of the minimal string. The proposed matrix model dual has a tree-level string equation
\beq
F(u) = \int_{\mathcal{C}} \frac{\d y}{2\pi \i} \, e^{2\pi y} \left( y-\sqrt{y^2-u-2 W(y)}\right),~~~W = \sum_i \lambda_i e^{-(2\pi - \alpha_i)y}.
\eeq
When $0\leq \alpha \leq \pi$ it reduces to the string equation for sharp defects. We verified some non-trivial limit of this formula in \cite{Turiaci:2020fjj}. For example, when $\alpha=2\pi$ we recover the JT gravity spectral curve. An interesting application of this formula can be found in the context of the discharge of near-extremal charged 4d black holes \cite{Brown:2024ajk}.

\medskip

More recently, with Eberhardt  \cite{Eberhardt:2023rzz} we showed that the Weil-Petersson measure on the moduli space with boundaries and defects is given by
\beq
\frac{\omega}{2\pi^2} = \kappa_1 + \sum_i \frac{b_i^2}{4\pi^2} \psi_i - \sum_j \frac{\alpha_j^2}{4\pi^2} \psi_i + \sum_{I\subset\{1,\ldots,m\}, \alpha_I \geq 0} \frac{\alpha_I^2}{4\pi^2}\delta_{0,I}
\eeq
where the last sum is over subsets of the defects and
\beq
\alpha_I = 2\pi - \sum_{i\in I} (2\pi - \alpha_i).
\eeq
Finally, $\delta_{g,I} \in \text{H}^2(\mathcal{M}_{g,n})$ are the Poincare duals of boundary divisors, which in physics language can be interpreted as a contact term when vertex operators collide. The boundary divisor separates a surface of genus $g$ with the punctures inside the set $I$. Their purpose is to precisely correct the issues with the Deligne-Mumford compactification. We gave some evidence that the GPI computed via this Weil-Petersson measure produces matches the matrix integral of \cite{Turiaci:2020fjj} in some concrete example. It is an open question to provide a proof of this duality.

\subsection{Fermionic JT gravity} 

We can change the theory in more subtle ways than a modification of the action. For example when summing over topologies we can keep track of spin structure, even if no dynamical fermions are present in the geometry. In this fashion, fermionic JT gravity can be defined, similarly to bosonic JT gravity, in terms of a $BF$ theory with group $\SL(2,\mathbb{R})$. The sign that we dropped in the bosonic theory encodes the information about the surfaces spin structure.

\medskip

One can define two different ways to sum over spin structures. First define the ``mod 2 index'' topological theory, considering closed surfaces first. Given a spin structure, we can study the Dirac operator $\slashed{D}$ and zero-modes come in pairs of opposite chirality. The number of say positive chirality zero-modes mod 2 is a topological invariant we call $\zeta$. Then we can define ``mod 2 index'' topological theory by an insertion of $(-1)^\zeta$ in the sum over spin structures. A nice discussion can be found in section 3 of \cite{Dijkgraaf:2018vnm} or section 2 of \cite{Stanford:2019vob}. If one wants to define it on a surface with a boundary one runs into an anomaly which precisely reproduces something we will see below in the matrix integral\footnote{A convenient way to compute the mod 2 index on surfaces with boundaries is by comparing different spin structures that are the same on $\partial \Sigma$. This is the relevant question anyways in the application to holography. The reason is that the mod 2 index is local so can remove the boundaries, glue them for example, and then evaluate it through Dirac operator zero-modes.}.

\medskip

\paragraph{Theory without mod 2 index} The fermionic JT gravity partition function $Z_{g,n}^{\rm ferm.}$ is simply the product of the bosonic JT gravity partition function $Z_{g,n}^{\rm bos.}$ on each surface of genus $g$ and $n$ boundaries multiplied by the sum over spin structures, namely
\beq
Z_{g,n}^{\rm ferm.}=  \Big(\sum_{\rm spin} 1\, \Big)\cdot Z_{g,n}^{\rm bos.} = 2^{2g+n-2}(1+(-1)^{n_R})Z_{g,n}^{\rm bos.}
\eeq
This is consistent with the fact that an orientable two-manifold with a spin structure has an even number of Ramond boundaries\footnote{A simple way to see this is the following. A three-holed sphere satisfies $\updelta_1\updelta_2\updelta_3 = -1$. The product of $\updelta_1\updelta_2\updelta_3$ over all three-holed spheres gives $\prod_{Y} \updelta_{Y_1} \updelta_{Y_2 } \updelta_{Y_3} = (-1)^\chi = (-1)^n$. Since all internal boundaries are counted twice, being shared by two three-holed spheres, then $\prod_{Y} \updelta_{Y_1} \updelta_{Y_2 } \updelta_{Y_3} = (-1)^{n_{\rm NS}}$. Therefore $(-1)^{n_{\rm NS}} = (-1)^{n-n_{\rm R}} = (-1)^n$ implying that $(-1)^{n_{\rm R}}=1$.}.

The partition function $Z_{g,n}^{\rm ferm.}$ can be reproduced by the following matrix integral. Consider a Hilbert space $\mathcal{H} = \mathcal{H}_b \oplus \mathcal{H}_f$ composed of bosonic and fermionic subspaces of dimensions $L_b$ and $L_f$, respectively. Assume with hindsight that $L_b=L_f=L/2$ where $L={\rm dim}\,\mathcal{H}$. Work in a basis where the fermion parity operator $(-1)^{\sf F}$ is represented as
\beq
(-1)^{\sf F} = \left(\begin{array}{c|c}
I & 0 \\
\hline 0 & -I
\end{array}\right).
\eeq
Consider a Hamiltonian of the form that commutes with fermion parity
\beq
H = \left(\begin{array}{c|c}
H_b & 0 \\
\hline 0 & H_f
\end{array}\right),
\eeq
where $H_b$ and $H_f$ are statistically independent matrices drawn from the GUE ensemble. The symmetry of the ensemble consists on change of basis compatible with the fermion parity operator, namely
$U = \left( \begin{array}{c|c}
     U_b & 0 \\
    \hline 0 & U_f
\end{array}\right)$
implying that $H_b \to U_b^{-1} H_b U_b$ and $H_f \to U_f^{-1} H_f U_f$. The measure over eigenvalues of $H_b$ and $H_f$ is therefore that of two independent GUE ensembles\footnote{Using the double-cone one can show this structure survives in higher dimensions \cite{Chen:2023mbc}.}. 

\medskip 

A boundary with NS structure with anti-periodic fermions corresponds to an insertion of 
\beq
{\rm Tr}_{\mathcal{H}}\, e^{-\beta H}= {\rm Tr}_{\mathcal{H}_b}\,\, e^{-\beta H_b}+{\rm Tr}_{\mathcal{H}_f}\,\, e^{-\beta H_f} ,
\eeq
while a boundary with R structure with periodic fermions corresponds to an insertion of
\beq
{\rm Tr}_{\mathcal{H}}\, (-1)^{\sf F}\, e^{-\beta H}={\rm Tr}_{\mathcal{H}_b}\,\, e^{-\beta H_b}-{\rm Tr}_{\mathcal{H}_f}\,\, e^{-\beta H_f} .
\eeq
Using that the random matrix ensemble above leads to statistically independent bosonic and femionic sectors, the correlators in the large $e^{S_0}$ limit are given by
\bea
\left\langle \prod_{i=1}^{n_{\rm NS}} Z_{\rm NS}(\beta_i) \,  \prod_{i=1}^{n_{\rm R}} Z_{\rm R}(\beta_i)\right\rangle_{g,n}^{\rm conn.}&=& \left\langle \prod_{i=1}^n{\rm Tr}_{\mathcal{H}_b}\,\, e^{-\beta H_b} \right\rangle + (-1)^{n_{\rm R}}\left\langle \prod_{i=1}^n{\rm Tr}_{\mathcal{H}_f}\,\, e^{-\beta H_f} \right\rangle,\nn
&=& (1+(-1)^{n_{\rm R}}) \cdot (\text{GUE correlator, $e^{S_0}\to e^{S_0}/2$}),\nn
&=& \underbrace{2^{2g+n-2}(1+(-1)^{n_{\rm R}})}_{\text{Sum over spin structures}} \times \underbrace{(\text{GUE correlator, $e^{S_0}$})}_{\text{Reproduced by JT gravity amplitudes}}\nn
\ea
In the first line we used the statistical independence combined with the fact that $(-1)^{\sf F}$ will insert a minus sign for each Ramond boundary. In the second line we used the fact that both $H_b$ and $H_f$ have the same dimension and therefore both contribute the same as the GUE ensemble with a dimension given by half the total Hilbert space dimension. This rescaling produces a prefactor times the GUE correlator with conventional normalization of $e^{S_0}$. The prefactor is precisely $\sum_{\rm spin} 1$ while the GUE correlator is given by the bosonic JT gravity path integral, completing the duality.

\medskip

\paragraph{Theory with mod 2 index} Next, consider the theory that does include the mod 2 anomaly in the bulk. The result of the GPI is now given by the bosonic JT gravity answer multiplied by 
\beq
\sum_{\rm spin} (-1)^\zeta = 2^{g+n-1} \delta_{n_R,0}.
\eeq
Notice that the power that appears here is not the Euler characteristic and therefore cannot be absorbed fully in a shift of $S_0$. (This is only true with boundaries! In closed surfaces $n=0$ and we can shift $e^{S_0}$ making the theory trivial.) 

\medskip

To reproduce this answer from a matrix integral we are forced to impose 
\beq
{\rm Tr}\, (-1)^{\sf F} e^{-\beta H} = 0,
\eeq
for each member of the ensemble (as opposed to on average, as in the previous case). Assume also that an insertion of a NS boundary corresponds to 
\beq
\sqrt{2} \, {\rm Tr} e^{-\beta H}
\eeq
Then the random matrix prediction is
\beq
\underbrace{(\sqrt{2})^{n} (\sqrt{2})^{2g+n-2}}_{2^{g+n-1} = \sum_{\rm spin} (-1)^\zeta}\times (\text{GUE correlator, $e^{S_0}$})
\eeq
The first factor comes from the $\sqrt{2}$ multiplying the trace. The second factor comes from a shift of $e^{S_0}$ to match the disk. The product of the two prefactors reproduces the sum over spin structures.

\medskip

We can interpret these rules as arising from a quantum-mechanical model with an anomaly in $(-1)^\F$. This arises for example in any quantum mechanical theory with an odd number of Majorana fermions. The Hilbert space is a representation of the Clifford algebra and there is no chirality in odd dimensions. From the path integral perspective there is an odd number of fermion zero-modes which cannot be soaked by interactions which would involve an even number of fermions. The factor of $\sqrt{2}$ relating path integral to traces can be inferred by considering $N$ free fermions (the path integral is $2^{N/2}$ while the dimension of the Hilbert space is $2^{(N-1)/2}$ if $N$ is odd.) These results are therefore quite natural from a holographic perspective.

\subsection{Unorientable JT gravity}

Next, we consider ``unorientable JT gravity'' where we sum over both orientable and unorientable surfaces. For simplicity we will first consider a theory where we discard, and not keep track of, spin structure. We will add this extra layer of complication in the following section.

\medskip 

In the $BF$ description, unorientable JT gravity can be obtained by replacing the gauge group $\PSL(2,\mathbb{R})$ by its double cover $\text{PGL}(2,\mathbb{R})$. This is the group of $2\times 2$ invertible real matrices modulo multiplication by a nonzero real scalar. This group includes the element 
\beq
U_{\sf R}=\lambda \left(\begin{array}{cc}
     1 &0  \\
     0 & -1
\end{array}\right),~~~~\lambda \in \mathbb{R},
\eeq
which $\PSL(2,\mathbb{R})$ would not include. Multiplication by a nonzero real scalar can be used to set the determinant of any matrix to be either $1$, meaning its in $\PSL(2,\mathbb{R})$, or $-1$, meaning its the product of a matrix in $\PSL(2,\mathbb{R})$ times $U_{\sf R}$. This double cover does not care about spin structure since $I$ and $-I$ are still identified in $\text{PGL}(2,\mathbb{R})$.

\medskip

What is the geometric interpretation of $U_{\sf R}$? Whenever a holonomy includes $U_{\sf R}$ the meaning is that there has been an orientation reversal. For example, in the case of the cylinder the ``holonomy'' from one boundary relative to the other could be $U_\varrho=\text{diag}(e^{\varrho/2},e^{-\varrho/2})$ (standard choice) or $ U_{\sf R} \cdot U_\varrho = \text{diag}(e^{\varrho/2},-e^{-\varrho/2})$ which involves an orientation reversal prior to gluing. The latter is referred to  as the twisted trumpet in \cite{Stanford:2019vob}. Therefore $Z_{0,2}$ in unorientable JT gravity is twice what we computed in section \ref{sec:2bdywh} receiving an equal contribution from both $U_\varrho$ and $U_{\sf R} \cdot U_\varrho$. 

\medskip 

There are two ways to sum over unorientable spaces. Unorientable surfaces can be made out of crosscaps, which are built by making a hole on a surface and closing it by gluing diametrically opposite points. The number of crosscaps on a surface $n_{\text{c}}$ is defined mod 2 and therefore when summing over unorientable surfaces we have the choice of adding or not a factor of $(-1)^{n_{\text{c}}}$. We will argue below that unorientable JT gravity is dual to either the GOE or the GSE ensemble depending on whether we add this sign or not. 

\medskip

Let us first collect some facts about the GOE and GSE ensembles. They correspond to systems with a time-reversal symmetry generated by an anti-unitary operator ${\sf T}$ that satisfies either ${\sf T}^2=1$ (GOE) or ${\sf T}^2=-1$ (GSE). A simple argument implies that all eigenvalues in the GSE ensemble are at least two-fold degenerate. The eigenvalue measure in these ensembles is
\beq
\int \prod_{i=1}^L\d\lambda_i \, \prod_{j<i} |\lambda_i -\lambda_j|^\upbeta \, e^{-L V(\lambda_i)},
\eeq
with $\upbeta=1$ ($\upbeta=4$) for the GOE (GSE) ensmeble and $\upbeta=2$ for GUE. $L$ is the size of the matrix prior to the double-scaling limit. In all cases the resolvent is defined via $R(x)= \sum_j (x-\lambda_j)^{-1}$. This is equalt to ${\rm Tr} (x-H)^{-1}$ for GUE and GOE, but for GSE the two-fold degeneracy implied by the algebra leads to ${\rm Tr} (x-H)^{-1} = 2\sum_j (x-\lambda_j)^{-1}$. For completeness we quote here the loop equations for these ensembles. They take the same form $2 x y(x) R_{g,n+1}(x,I)+x F_{g,n+1}(x,I) \sim 0$ where
\bea
F_{g,n+1}(x,I) &=& \Big(1-\frac{2}{\upbeta}\Big) \partial_x R_{g-\frac{1}{2},n+1}(x,I)\nonumber\\
&&+ R_{g-1,n+2}(x,x,I) + \sum_{\rm stable} R_{h_1,|I_1|+1}(x,I_1) R_{h_2,|I_2|+1}(x,I_2)\nonumber\\
&&+ 2 \sum_{k=1}^n \Big( R_{0,2}(x,x_k) + \frac{1}{\upbeta} \frac{1}{(x-x_k)^2}\Big) R_{g,n}(x,I/x_k).\label{eq:LEGB}
\ea
The crucial new ingredient here is the term in the first line which is non-vanishing when $\upbeta=1$ or $4$ (and vanishes for the GUE ensemble with $\upbeta=2$). This term produces terms in the topological expansion with odd $\chi$. It also affects the loop equations in a way that $R_{g,n+1}$ is not given any longer by a residue evaluated at the end-points and one needs to evaluate the full integral around the cut.

\medskip

It is convenient to relate the GOE and the GSE ensemble loop equations so we can focus on one of them when comparing with gravity. It is a useful exercise to check using \eqref{eq:LEGB} that the loop equations for the GSE ensemble, written in terms of
\beq
\widetilde{R}_{g,n}(x_1,\ldots,x_n;L) = 2^{n} R_{g,n} (x_1,\ldots,x_n;L/2),
\eeq
are identical to the GOE loop equations although with the sign of the crosscap (terms with odd $\chi$) reversed. This matches with the gravity expectation for the following two reasons. First, $\widetilde{R}_{g,n}$ corresponds to the trace in the GSE ensemble, since the factor of $2^{n}$ precisely accounts for the two-fold degeneracy. The rescaling $L\to L/2$ amounts to a simple shift of $S_0$ that is required for both ensembles to produce the same disk partition function. Therefore, in these conventions, the GOE partition function is identical order by order to the GSE partition function, except for a change of sign of crosscap contributions, as expected from the insertion of $(-1)^{n_{\text{c}}}$ in gravity. We then focus on the GOE case for concreteness.

\medskip

Let us describe now the gravity calculation, starting with  orientable surfaces. Since we are gauging orientation reversal, the orientation of a given boundary is not meaningful. But once we choose an orientation in one boundary, the relative orientation of the other $n-1$ boundaries will be meaningful since they can be measure by ``holonomies'' obtained from parallel transport between boundaries. This generates $2^{n-1}$ topologically distinct contributions. The partition function in any of them is exactly the same as the one we computed in oriented (bosonic) JT gravity. This therefore produces a simple factor
\beq
\text{(Path integral over oriented surface)} = 2^{n-1} \cdot Z^{\rm bos.}_{g,n},
\eeq
and reproduces the factor of two we argued above for the double-trumpet. It is easy to check that, for genus zero, the loop equations of the GOE and GUE resolvents are related by
\beq
R_{0,n}^{\rm GOE}(x_1,\ldots,x_n) =2^{n-1} R_{0,n}^{\rm GUE}(x_1,\ldots,x_n) ,
\eeq
matching the gravity expectation. 

\medskip

Let us now consider unorientable surfaces. While orientable surfaces are made out of three-holed spheres, to build a hyperbolic metric on an arbitrary possibly unorientable we need another kind of building block. This is obtained from a three-holed sphere by replacing one or more of its boundary circles with crosscaps. If $\chi$ is even we can make unorientable surfaces only out of orientation-reversal when gluing. How to compute the path integral on unorientable surfaces? At this point we are forced to use the torsion; JT gravity is not tree-level exact anymore on unorientable surfaces!

\medskip

To complete the calculation we need to evaluate the torsion of a cross-cap. We can represent it by a cylinder with two circle boundaries $S$ and $S'$. On $S'$ we identify antipodal points making it an internal circle now. If we denote by $U=\text{diag}(e^{b/2},e^{-b/2})$ the holonomy around $S$ then the holonomy around $S'$, which we call $V$, should satisfy $U=V^2$. Since going around $S'$ involves an orientation reversal we take $V= U_{\sf R}\cdot \text{diag}(e^{b/4},e^{-b/4})= \text{diag}(e^{b/4},-e^{-b/4})$. By an argument similar to the one that applied to the circle, the torsion of the crosscap involves the determinant of the operator $s \to V s V^{-1} - s$, with eigenvalues $-1 - e^{\pm b/2}$ and a careful treatment of the zero-modes. The result is
\beq
\tau_C = 2 \cosh^2 \frac{b}{4} \cdot \d b \cdot (\d \varrho)^{-1}.
\eeq

We can now derive a measure for integration over the moduli space of hyperbolic surfaces which are possibly unorientable. They can be built out of three-holed spheres with orientation reversing gluing (this would be the trivial case) but they can also involve three-holed-sphere where either one or two boundaries are replaced by crosscaps. Using the gluing rules described in the previous lecture one automatically gets 
\beq
\tau_Y = \prod_{i=1}^{3g-3} \d b_i \, \d\varrho_i \, \prod_{\alpha=1}^{n} \frac{1}{2} \coth \frac{b_\alpha}{4}\, \d b_\alpha.
\eeq
Here $b_\alpha$ are the lengths of the crosscap geodesics. The twist cancels in the gluing, as it should. Precisely this measure was previously ``bootstrapped'' by Norbury \cite{norbury2007lengthsgeodesicsnonorientablehyperbolic}, with the assumption that it takes a factorized form and that its independent of choice of coordinates. This is consistent with the gravity calculation which should be invariant under change of coordinates (the factorized form is not obvious although very reasonable given the gluing property of the torsion).

\medskip

We can perform another small check; compute the crosscap with one boundary. Recalling the trumpet with torsion boundary conditions, and the gluing relations for the torsion, leads to
\beq
Z_{\frac{1}{2},1}(\beta) = \int_0^\infty \d b \, Z_{\rm JT}^{\rm trumpet}(\beta ,b) \, \frac{1}{2} \coth \Big(\frac{b}{4} \Big).
\eeq
Even though this is divergent, the integral can be successfully compared to the prediction from loop equations arising from the first term on the right hand side of \eqref{eq:LEGB}. This was actually shown to all orders recently by Stanford using a prescription that regularizes the integrals \cite{Stanford:2023dtm}.

\subsection{Unorientable fermionic JT gravity}

We can combine the last two generalizations and sum over both orientable and unorientable surfaces, and over their spin structures. This case is quite involved so we will summarize some facts to orient the reader, referring to the original work in \cite{Stanford:2019vob} and references therein for a more thorough treatment. In the $BF$ description we included orientable surfaces by replacing $\PSL(2,\mathbb{R})$ by $\text{PGL}(2,\mathbb{R})$, and we included a spin structure by replacing $\PSL(2,\mathbb{R})$ by $\SL(2,\mathbb{R})$. To do both things we need to take the appropriate double cover of $\SL(2,\mathbb{R})$. There are two such groups which are called $\pin^+$ and $\pin^-$:

\medskip 

For $\pin^+$ one wants the group of real matrices with determinant $\pm 1$ meaning one again represents orientation-reversal ${\sf R}$ by $\text{diag}(1,-1)$. This means that ${\sf R}^2=1$. The CPT theorem implies that any field theory on a Lorentzian spacetime has a non-unitary symmetry ${\sf RT}$. Although this symmetry has universal properties, the definition of ${\sf R}$ and ${\sf T}$ separately depends on choices. If ${\sf R}^2 =1$ the CPT theorem implies a time-reversal symmetry exists such that ${\sf T}^2 = (-1)^\F$, see section 5 of \cite{Witten:2025ayw} for a very clear explanation.\footnote{\textbf{v2:} In a previous version of these notes, it was incorrectly stated that ${\sf RT}^2 = (-1)^{\sf F}$. I thank Jan Boruch and Elisa Tabor for pointing this out.}

\medskip

For $\pin^-$ we work with the group of matrices of the determinant $1$ that are real or imaginary, which implies that one includes the element ${\sf R} = \text{diag}(\i , -\i )$. In this case ${\sf R}^2=-1$ which acts trivially on bosonic fields and produces a negative sign on fermions. This implies that it acts as ${\sf R}^2 = (-1)^\F$. The CPT theorem now implies that time reversal acts on the boundary as ${\sf T}^2 = 1$.

\medskip

The procedure should now be clear. For a given orientation and spin structure, we can perform the JT gravity path integral obtaining the result in the previous lecture. The nontrivial ingredient now is the possibility to incorporate a sum over $\pin^\pm$ structures together with possible bulk topological theories that weight them in different ways. We conclude by summarizing some aspects of these possibilities:

\medskip

\paragraph{Sum over $\pin^-$ structures} The topological invariant on a surfaces with ${\rm pin}^-$ structure that generalizes the mod 2 index is given by
\beq
\exp(-\i \pi \eta/2), 
\eeq
where $\eta$ is the Atiyah-Patodi-Singer eta invariant of the self-adjoint operator $\i \slashed{D}$. For any manifold, this phase is the eighth-root of unity giving essentially $8$ possible theories where
\beq
Z^{\rm unor.~ferm.}_{g,n}=Z^{\rm bos.}_{g,n} \cdot \sum_{{\rm pin}^-} \exp(-\i \pi N\eta/2), 
\eeq
and $N$ is defined mod 8. On orientable surfaces the eta invariant reduces to the mod 2 invariant and we recover the two theories analyzed earlier for fermionic JT gravity. Sums over pin structures more generally can be computed by combining the relation to the mod 2 index with the locality of the eta-invariant, together with the value of $\eta$ for a crosscap.

\medskip

Let us work out one example to illustrate how the matrix integral duality works. Consider the topological theory with $N=0,4~$mod 8. One can show that
\bea
\sum_{{\rm pin}^-} \exp(-\i \pi 0\eta/2) &=& 2^{2g+n-2} (1+(-1)^{n_{\rm R}}), \\
\sum_{{\rm pin}^-} \exp(-\i \pi 4\eta/2) &=&  2^{2g+n-2} (-1)^{n_{\rm c}} (1+(-1)^{n_{\rm R}}).
\ea
Multiplying this by the unoriented JT gravity partition function, which we already determined is dual to the GOE/GSE ensemble, leads to the two matrix ensembles 
\bea
&&N=0~\text{mod}~8 ~~~\Rightarrow~~~H=\left(\begin{array}{cc}
\text{GOE}_1 & 0 \\
0 & \text{GOE}_2
\end{array}\right),\\
&&N=4~\text{mod}~8 ~~~\Rightarrow~~~H=\left(\begin{array}{cc}
\text{GSE}_1 & 0 \\
0 & \text{GSE}_2
\end{array}\right).
\ea
The reason is simple; the presence of the term $(1+(-1)^{n_{\rm R}})$ means that fermion parity is a symmetry of the ensemble, since this term reproduces the sum over statistically-independent sectors. The fact that they differ by a factor of $(-1)^{n_{\rm c}}$ implies that the ensembles in each sector are either GOE or GSE depending on the case. Purely in terms of symmetry, the case $N=0~$mod 8 corresponds to a theory where quantum mechanically there is a $\mathbb{Z}_2^\T \times \mathbb{Z}_2^\F$ group generated by $\T^2 =1$ and $(-1)^\F$. When $N=4$ mod 8 the only difference is that now $\T^2=-1$. Similarly, all choices of $N$ can be mapped to possible anomalies in the classical $\mathbb{Z}_2^\T \times \mathbb{Z}_2^\F$ symmetry.

\medskip

\paragraph{Sum over $\pin^+$ structures} The only invariant we can define on unorientable ${\rm pin}^+$ structures is a mod 2 index of the Dirac operator (non-chiral since we include unorientable surfaces). This is sometimes denoted by
\beq
(-1)^{\widetilde{\zeta}}.
\eeq
Therefore there are two theories corresponding to $\pin^+$ structures. When we sum over pin structures without $(-1)^{\widetilde{\zeta}}$, this corresponds to a non-anomalous boundary theory with both $\sf{T}$ and $(-1)^{\sf F}$ but
\beq\label{eq:t2mfbbb}
{\sf T}^2 = (-1)^{\sf F}.
\eeq
making the group $\mathbb{Z}_4^{\sf T}$ where the superscript indicates it involves time reversal. This means that the Hilbert space decomposes into two sectors. The bosonic sector leads to the GOE ensemble and the fermionic sector to the GSE ensemble. This can be verified using the result
\beq
\sum_{{\rm pin}^+} 1 = 2^{2g+n-2}(1+(-1)^{n_{\rm R}+n_{\rm c}}).
\eeq
The term $(1+(-1)^{n_{\rm R}+n_{\rm c}})$ suggests the presence of two sectors while the relative $(-1)^{n_{\rm c}}$ suggests that ${\sf T}^2$ has a different sign between them, consistent with \eqref{eq:t2mfbbb}. In the presence of the mod 2 topological theory, we expect an anomaly. There is only one possibility which can be written as
\beq
{\sf T}^2 = \i (-1)^{\sf F}.
\eeq
The anomaly can be moved around (for example can be put in the fermion parity which would then square to minus one) but cannot be removed. Importantly, the antiunitarity of ${\sf T}$ implies 
\beq
{\sf T} (-1)^{\sf F}= - (-1)^{\sf F} {\sf T}.
\eeq
Therefore time-reversal exchanges the bosonic and fermionic blocks and does nothing else. This is consistent with another result
\beq
\sum_{{\rm pin}^+} (-1)^{\widetilde{\zeta}} = 2^{2g+n-1} \,\delta_{n_{\rm R},0}\,\delta_{n_{\rm c},0}.
\eeq
We see the contribution with cross-caps vanishes identically after summing over pin structures, consistent with having two identically distributed blocks in the GUE ensemble. The reader can find in section 2 of \cite{Stanford:2019vob} an explanation on the origin of some of these identities on the sum over pin structures.

\section{Supergravity}\label{sec:sugra}
In this final section we present some results on JT supergravity and its dual random matrix ensemble. As we shall see, these theories force us to combine what we have learned so far about generalizations that involve spin structure, global symmetries of the boundary theory, and the possibility of bulk topological theories. We will mention some physics motivations to considering these theories as we go along. 
\subsection{$\mathcal{N}=1$ JT gravity}
\subsubsection{Basics}
A definition of the theory using superspace can be found in \cite{Forste:2017kwy}. As should be clear by now, for our purposes, the most efficient way to describe a generalization of JT gravity with $\mathcal{N}=1$ is through the $BF$ description \cite{Grumiller:2017qao}. We replace the group $\SL(2,\mathbb{R})$ by its smallest supersymmetric generalization
\beq
\SL(2,\mathbb{R}) \to {\rm OSp}'(1|2)={\rm OSp}(1|2)/\mathbb{Z}_2.
\eeq
This is the group of linear transformations of two bosonic variables $u,v$ and one fermionic $\uptheta$ that preserves the symplectic form
\beq
\omega = \d u \, \d v + \frac{1}{2} \d \uptheta^2.
\eeq
We are modding out by the transformation $u,v|\uptheta \to -u,-v|-\uptheta$ which commutes with everything. The bosonic generators can be written as 
\beq
{\sf e} = \left(\begin{array}{cc|c}
     0 &1  &  0  \\
     0 &0  & 0\\
     \hline
     0&0& 0
\end{array}\right), ~~~{\sf f} =  \left(\begin{array}{cc|c}
     0 &0  &  0  \\
     1 &0  & 0\\
     \hline
     0&0& 0
\end{array}\right),~~~~{\sf h} = \left(\begin{array}{cc|c}
     1 &0  &  0  \\
     0 &-1  & 0\\
     \hline
     0&0& 0
\end{array}\right),
\eeq
and the fermionic ones as
\beq
{\sf q}_1 =  \left(\begin{array}{cc|c}
     0 &0  &  1  \\
     0 &0  & 0\\
     \hline
     0&-1& 0
\end{array}\right), ~~~{\sf q}_2 = \left(\begin{array}{cc|c}
     0 &0  &  0  \\
     0 &0  & 1\\
     \hline
     1&0& 0
\end{array}\right). 
\eeq

\medskip

\textbf{Exercise:} Write explicitly the action of $\mathcal{N}=1$ JT supergravity using the $BF$ formulation. The bosonic generators give rise to the dilaton and metric, while the fermionic generators give rise to the dilatino and gravitino. In order to do this, remember to replace in all formulas the trace over the matrices by the supertrace, which for a given supermatrix $M=\left( \begin{array}{c|c}
     A &B  \\
     \hline C & D
\end{array}\right)$ is given by ${\rm STr}\, M = {\rm Tr}\, A - {\rm Tr} \,D$. Moreover, whenever a determinant of a matrix appeared before we should replace it by the Berezinian which if $B$ or $C$ blocks vanish is ${\rm Ber}(M) = {\rm det}(A)/{\rm det}(D)$ (this will be important later but irrelevant for the action calculation).

\medskip

The JT gravity path integral will naturally continue to localize into flat connections, modulo gauge transformations. This space again has multiple components and the connection with gravity forces us to restrict to one where all holonomies are hyperbolic (after reducing modulo odd variables). We can refer to this as a $\mathcal{N}=1$ generalization of Teichmuller space \cite{1987CMaPh.113..601C,1990CMaPh.135...91D}.

\medskip

When used to described hyperbolic surfaces, the geometric information is encoded in the $BF$ formulation holonomies in the following way
\beq
U = \pm \left(\begin{array}{cc|c}
     e^{b/2} & 0 & 0 \\
     0 &  e^{-b/2} & 0\\
     \hline
     0 & 0 & \updelta 
\end{array}\right),
\eeq
where $b$ is the geodesic length and  $\updelta = \pm 1$ denote the spin structure. The overall sign is not meaningful in $\text{OSp}'(1|2)$.

\medskip

The first step, as we did for JT gravity, is to study the disk and the cylinder. The path integral over the disk reduces to a straightforward generalization of the Schwarzian action with one supercharge, defined in \cite{Fu:2016vas}. The boundary mode is parameterized by super-reparametrizations of the $\mathcal{N}=1$ super-line $(\tau,\theta)$ that includes both a bosonic component $\tau\to f(\tau)$ as well as fermionic $\theta \to \theta + \eta(\tau)$. The action is
\beq
I = - \Phi_r \int \d \tau \, \{ \tan \frac{\pi f}{\beta}, \tau\} + \eta\, \eta''' + 3 \eta' \eta'' - \{ \tan \frac{\pi f}{\beta} , \tau\} \eta \, \eta'.
\eeq
The zero-mode of this action are the generators of the isometries of the $\mathcal{N}=1$ hyperbolic disk. This isometry group is precisely ${\rm OSp}(1|2)$ of dimension $3|2$. The fermion zero-modes have a behavior $\eta \sim  e^{ \pm \i \frac{1}{2} \frac{2\pi \tau}{\beta}}$.

\medskip

 The partition function on the disk in the R sector (periodic fermions along time direction) vanishes since periodic fermions cannot be smoothly extended given that time is contractible at the horizon. The path integral for the NS sector (antiperiodic fermions along time direction) localizes by a supersymmetric analog of the Duistermaat-Heckman formula \cite{Stanford:2017thb} and the one-loop determinant is again given by the ``rotation angles'' of a $\U(1)$ symmetry. The one-loop determinant for bosons and fermions are
\beq
Z_{\rm one-loop} = \underbrace{\prod_{n\geq 2} \frac{1}{n/(2\beta)}}_{\text{Schwarzian mode}} \,\, \cdot \,\underbrace{ \prod_{m\geq 3/2} m/(2\beta) }_{\text{Fermion}} = \sqrt{\frac{2}{\pi \beta}}.
\eeq
To derive this result we can use that in zeta-function regularization $\prod_{m\geq 1/2} m/(2\beta) = \sqrt{2}$, reproducing the partition function of a single Majorana fermion. The final answer for the disk partition function is
\beq
Z_{\rm disk} = \sqrt{\frac{2}{\pi \beta}} \, e^{S_0 +\frac{\pi^2}{\beta}}.
\eeq
The different factor of $\beta$ compared to its bosonic counterpart can be traced back to the fact that the $\mathcal{N}=1$ has two new fermionic isometries, which induces a new factor of $\beta$ in the numerator. The spectral curve derived from this formula is
\beq
\rho(E) =e^{S_0} \frac{\sqrt{2} \cosh( 2\pi \sqrt{E})}{\pi \sqrt{E}},~~~~~y(x) = - \frac{\sqrt{2} \cos(2\pi \sqrt{-x})}{\sqrt{-x}}.
\eeq
It might be surprising that the density of states has an inverse square-root edge. We will see that this is the natural behavior for a matrix ensemble with $\mathcal{N}=1$ supersymmetry.

\medskip 

A similar calculation on the trumpet with antiperiodic fermions gives
\beq
Z_{\rm SJT}^{\rm trumpet} (\beta, b) = \frac{1}{\sqrt{2 \pi \beta}} \, e^{-\frac{b^2}{4\beta}}.
\eeq
This has the same power of $\beta$ since the trumpet has no fermionic isometries\footnote{One can give a simple $BF$ argument for this. The holonomy around the hole is $\pm {\rm diag}(e^{b/2},e^{-b/2}, \updelta)$. There are no fermionic matrices that commute with this holonomy for  $b>0$.}. It is multiplied by $\sqrt{2}$ compared to the bosonic answer, the partition function of a Majorana fermion. The cylinder partition function is
\beq
Z_{0,2}(\beta_1,\beta_2) = 2 \int b \d b \, Z_{\rm SJT}^{\rm trumpet}(\beta_1,b) Z_{\rm SJT}^{\rm trumpet}(\beta_2,b).
\eeq
The factor of $2$ arises from the two spin structures ``orthogonal'' to the gluing geodesic, described by a twist holonomy ${\rm diag}(e^{\varrho/2},e^{-\varrho/2},\pm1)$. Overall the answer is four times the GUE one. This suggests that eigenvalues come with a two-fold degeneracy as expected between bosonic/fermionic states in a supersymmetric theory.

\medskip

Since the time circle is non-contractible on the trumpet we can evaluate the partition function with periodic fermions. The answer still vanishes, although now due to the presence of a fermion zero-mode living near each boundary. Therefore the partition function with any number of Ramond boundaries vanishes identically. This implies that the Witten index of the theories in the ensembles does not fluctuate.

\subsubsection{Measure over moduli space}

For other topologies, we can compute the measure over moduli space of $\mathcal{N}=1$ hyperbolic surfaces. Again, we have either a symplectic or torsion approach available. Both are applicable but the symplectic approach is not particularly simpler and does not apply on unorientable surfaces; for this reason we will use the torsion again. 

\medskip

To compute the combinatorial torsion we need to determine a supergroup measure. We did that earlier by starting with a nondegenerate quadratic form, from which we derived a measure involving a determinant. The generalization of this to supergroups is the following. First, the quadratic form should involve a supertrace instead of a trace to guarantee its conjugation invariance. Second, we need a generalization of $|{\rm det}\, M|$ for supergroups to determine a measure. The determinant is replace by the Berezinian while the absolute value is replaced by ${\rm Ber}'\, M  = {\rm sgn}({\rm det}\,A) \, {\rm Ber}\, M$. The torsion associated to bosonic groups is positive, given that it involves positive measures. This is not the case for supergroups and there is some arbitrariness in assigning signs. Even more, on unorientable surfaces the path integral can even be complex.

\medskip

\paragraph{Torsion of a circle} Let us begin considering the case of a circle since it is simpler. Take its holonomy to be $U=\pm {\rm diag}(e^{b/2},e^{-b/2}, \updelta)$. The torsion is the Berezinian of the operator $\partial s = U s U^{-1} - s$, excluding zero-modes. A simple calculation leads to the bosonic eigenvalues $e^{\pm b} -1$ and fermionic $\updelta e^{\pm b/2}-1$. The zero-modes contribute $\d b \cdot (\d \varrho)^{-1}$ for the same reason as in the bosonic theory. The final answer is
\beq
\tau_S = - \updelta (e^{b/4}+ \updelta e^{-b/4})^2 \d b \cdot (\d \varrho)^{-1} = \begin{cases}
    4 \sinh^2(\frac{b}{4}) \,\d b \cdot (\d \varrho)^{-1} ~~~~\text{NS spin structure,}\\
    - 4 \cosh^2 (\frac{b}{2}) \,\d b \cdot (\d \varrho)^{-1} ~~~\text{R spin structure.}
\end{cases}
\eeq

\medskip

\paragraph{Torsion of a three-holed sphere} The description of the moduli space can be made very similar to the case of bosonic gravity. We have three holes with holonomy $U$, $V$, $W$ constrained by $UVW=1$ and defined only modulo overall conjugation. This leaves total of $3|2$ moduli. The three bosonic ones are the usual geodesic lengths while the remaining 2 moduli are fermionic.

\medskip 

\noindent Take $x\in {\rm osp}(1|2)$ and consider the quadratic form $|| x ||^2 = 2 \,{\rm STr}\, x^2= 4 x_h^2 + 4 x_e x_f + 8 x_1 x_2$. The Berezinian of this metric is one, so the natural measure is $[\d x_e \, \d x_f \, \d x_h | \d x_1 \, \d x_2]$. The normalization of the quadratic form is chosen so that it reduces to the one considered in the bosonic case, which we show to coincide with the Weil-Petersson form with the usual normalization. Finally the form on the group manifold in terms of $U$ is the measure on $U^{-1} \d U$.

\medskip

\noindent We parametrize the holonomies by $U= R U_0 R^{-1}$ and $V= R V_0 R^{-1}$ with
\beq
U_0 = \updelta_1 \left( \begin{array}{cc|c}
    e^{b_1/2} & \kappa & 0 \\
     0 & e^{-b_1/2} &0 \\
  \hline 0 & 0 &\updelta_1
\end{array}\right)\cdot e^{\xi {\sf q}_1},~~~~~V_0 = \updelta_2 \left( \begin{array}{cc|c}
     e^{-b_2/2} & 0&0  \\
     1 & e^{b_2/2}  & 0 \\
   \hline 0 &  0 &\updelta_2
\end{array}\right)\cdot e^{\psi {\sf q}_2}.
\eeq
We can compute the torsion through $\tau = {\rm vol}(U)\, {\rm vol}(V) / {\rm vol}(R)$. We also need to find an equation that relates $\kappa$ with $b_3$. The final answer for the torsion is
\beq
\tau_Y = \frac{2 \sinh \frac{b_1}{2} \sinh \frac{b_2}{2} \sinh \frac{b_3}{2}}{(\updelta_1 e^{b_1/2}-1)(\updelta_2 e^{b_2/2}-1)}\, [ \d b_1\, \d b_2\, \d b_3 | \d \xi \d \psi]
\eeq
\textbf{Exercise:} Reproduce the torsion of $\mathcal{N}=1$ supergravity. This expression already implies that the Weil-Petersson volume on the three-holed sphere vanishes $V_{3,0}=0$, due to the presence of the fermionic zero-modes $\xi$ and $\psi$.

\paragraph{Measure over moduli space} We now can glue the pieces. Consider a closed surface first. We take the product of the torsion of all $2g-2$ three-holed spheres $t\in Y$ and the product of the inverse torsion of all circles $s\in S$. Since the torsion of the three-holed sphere is not manifestly invariant under permutation of boundaries we need to make an arbitrary choice of labels which we call $(b_1,b_2,b_3) \to (a_t,b_t,c_t)$ where $t\in Y$ is the three-holed sphere under consideration. The final answer after combining all terms is 
\beq
\tau = \frac{1}{2} (-1)^{w_{\rm R}} \prod_{s\in S} [\d b_s \d \varrho_s] \, \prod_{t\in Y} \frac{1}{4} \updelta_{a_t} \updelta_{b_t} e^{-(a_t+b_t)/4} (e^{c_t/4} - \updelta_{c_t} e^{-c_t/4}) [\d \xi_t \, \d \psi_t].
\eeq
The prefactor of $1/2$ comes from taking into account the $\mathbb{Z}_2$ symmetry $(-1)^{\sf F}$. $w_{\rm R}$ is the number of interior Ramond boundaries. Since this coincides with the symplectic measure the path integral of $\mathcal{N}=1$ JT gravity on closed surfaces is 
\beq
Z_{\rm SJT}{}_{g,0} = V_{g,0},
\eeq
where the volumes in this section corresponds to those of the moduli space of  $\mathcal{N}=1$ hyperbolic surfaces.

\medskip

A similar formula can be written in the case of surfaces with boundaries in an obvious way, although the trigonometric factors are not that nice. The final expression for the partition function with all NS boundaries, since otherwise it vanishes, is
\beq
Z_{g,n}(\beta_1,\ldots,\beta_n) = \prod_{e=1}^n \int_0^\infty b_e \, \d b_e\, Z_{\rm SJT}^{\rm trumpet}(\beta_e,b_e) \underbrace{V_{g,n}(b_1,\ldots,b_n)}_{\text{Volumes of $\mathcal{N}=1$ moduli space}}
\eeq
Finally, besides integrating over lengths and twist we also need to sum over spin structures. When performing this sum we are free to insert our mod 2 index topological theory $(-1)^\zeta$. This leads to two different $\mathcal{N}=1$ supergravity theories.

\medskip 

Some interesting properties are the following. First, one can show that $V_{g=0,n}=0$ due to the presence of fermionic moduli. Second, without an insertion of $(-1)^\zeta$ all volumes vanish! In other words the volume restricted to even spin structure is equal and opposite in sign to the volume restricted to odd spin structures. When we insert the mod 2 index the volumes are non-trivial.

\subsubsection{Duality with random matrices}
Classically $\mathcal{N}=1$ theories have a supercharge $Q$ and a fermion parity symmetry $(-1)^{\sf F}$. We get two classes of ensembles depending on whether the second symmetry survives quantization or not.

\medskip 

\paragraph{Theory with anomalous $(-1)^{\sf F}$} In this case $H=Q^2$ and $Q$ is a self-adjoint operator with no further structure since the fermion parity operator is anomalous. Therefore it is reasonable to expect $Q$ to be taken from the GUE ensemble, with the partition function being 
\beq
Z_{\rm gravity} (\beta) = \sqrt{2}\,{\rm Tr} \, e^{- \beta Q^2}.
\eeq
The spectral curve of the GUE ensemble for $Q$ is therefore 
\beq
Z_{\rm disk}(\beta) = \sqrt{2} \int_{-\infty}^{\infty} \d q \, \rho_0(q) \, e^{-\beta q^2},~~~~\rho_0(q) = \frac{\cosh(2\pi q)}{\pi}.
\eeq
This is a new situation; the spectral curve has support on the whole real axis. This is a double-scaling limit where both ends go to infinity. This is also in the same universality class as the Gross-Witten-Wadia model\footnote{This was important to resolve some puzzle when considering deformations of supergravity \cite{Rosso:2021orf}.}, and some non-perturbative aspects of this theory were studied by C. Johnson, F. Rosso and A. Svesko in \cite{Johnson:2021owr}.

\medskip

Since the model has a spectral curve with support on whole real axis the loop equations predict all higher genus corrections (other than disk and cylinder) to vanish. The reason is that for the GUE ensemble the loop equations get contributions solely from poles at the edges but there are no edges here. This is compatible with the result above for the theory where we sum over spin structures without $(-1)^\zeta$.

\paragraph{Theory with $(-1)^{\sf F}$ symmetry} In this case the Hilbert space should decompose into two sectors of fermionic and bosonic states $\mathcal{H}=\mathcal{H}_b \oplus \mathcal{H}_f$ with $L_{b/f}={\rm dim}(\mathcal{H}_{b/f})$. This implies that the supercharge now has the following structure
\beq
Q = \left( \begin{array}{c|c}
     0 & {\sf Q} \\
 \hline  {\sf Q}^\dagger  & 0
\end{array}\right),~~~~~H = \left( \begin{array}{c|c}
    {\sf Q} {\sf Q}^\dagger  & 0 \\
\hline    0 &  {\sf Q}^\dagger {\sf Q}
\end{array}\right).
\eeq
where ${\sf Q}:\mathcal{H}_f \to \mathcal{H}_b$ and ${\sf Q}^\dagger:\mathcal{H}_b \to \mathcal{H}_f$ is its adjoint. The Witten index of the theory is temperature-independent due to supersymmetry and it is given by $L_b-L_f$. The most general transformation that preserves this structure acts with an unitary $U = \left( \begin{array}{c|c}
     U_b & 0 \\
    \hline 0 & U_f
\end{array}\right)$, and the ensemble for the complex matrix ${\sf Q}$ should be invariant under
\beq\nonumber
{\sf Q} \to U_b^\dagger {\sf Q} U_f.
\eeq
We can use $U_b$ and $U_f$ to put ${\sf Q}$ in a canonical form, which if $L_b=L_f$ it takes the form
    \beq\nonumber
{\sf Q} = U_b^\dagger \, \begin{pmatrix}\lambda_1 & &&& \cr & \lambda_2 &&& \cr && \ddots && \cr
         &&&&\lambda_L \end{pmatrix} \, U_f
    \eeq
    with $\lambda_j \geq 0$. This is the singular value decomposition and it is applicable even if the matrix is rectangular (non-vanishing index). Since we know that the partition function with periodic fermions vanishes identically for all members of the ensemble, ${\rm dim} \mathcal{H}_b = {\rm dim} \mathcal{H}_f $ and the matrix should be square. 
This means that for each eigenstate in the fermionic sector there is an equivalent eigenstate in the bosonic sector, since each singular value will appear in both ${\sf Q}^\dagger {\sf Q}$ as well as ${\sf Q} {\sf Q}^\dagger$. This is also true if for some $j$ the eigenvalue accidentally vanishes $\lambda_j=0$. The ensemble above is one of the ten Altland-Zirnbauer ensembles \cite{Altland:1997zz}. 

After a simple calculation similar to the one for GUE, the measure of integration over the singular values becomes
\beq\nonumber
\int\d Q \, e^{- L{\rm Tr} V(Q^2)} \to \prod_{i} \int \d \lambda_i\,\lambda_i^\upalpha \, \prod_{i<j} |\lambda_i^2-\lambda_j^2|^{\upbeta}\,\, e^{-2L \sum_j V(\lambda_j^2)},
\eeq
where for our specific case of the complex matrix ${\sf Q}$, the coefficients are $(\upalpha,\upbeta)=(1,2)$. Although this ensemble was derived in the past \cite{Morris:1990cq,VZ,Anderson:1990nw,Anderson:1991ku,Dalley:1991qg,Dalley:1991vr}, it was interpreted for the first time by Stanford and Witten in \cite{Stanford:2019vob} as the ensemble appropriate to a many-body chaotic theory with one supercharge. These models were analyzed and solved non-perturbatively in \cite{Johnson:2020heh,Johnson:2020exp}. 

\medskip

We found that the index of $\mathcal{N}=1$ JT supergravity vanishes. What if we consider an ensemble with non-vanishing index? Since the index is independent of temperature, its value at $\beta=0$ computes $L_b-L_f$. This means that ${\sf Q}$ is now instead rectangular with a canonical form
$$
{\sf Q}=U_b^\dagger\begin{pmatrix}\lambda_1 & &&& \cr & \lambda_2 &&& \cr && \ddots && \cr
         &&&&\lambda_{L_f}\cr 0 &&&&\cr &0&&&\cr &&\ddots && \end{pmatrix}U_f 
$$
This guarantees that while the eigenvalues of ${\sf Q}^\dagger {\sf Q}$ are the $\{\lambda_j\}$, the eigenvalues of ${\sf Q}{\sf Q}^\dagger$ come also with $L_b-L_f$ identically zero eigenvalues. These are the BPS states forced by supersymmetry, while there can be accidentally BPS states corresponding to some $j$ for which $\lambda_j=0$. (We are assuming here that $L_b>L_f$ but similar statements can be made if $L_f > L_b$.) In terms of the measure over singular values, the only effect of the BPS states is to shift $\upalpha \to \upalpha + 2 |L_b-L_f|$. This will be important when we discuss $\mathcal{N}=2$ supergravity in the next section.

\medskip 
Based on these considerations, we can reproduce $\mathcal{N}=1$ supergravity by choosing a Hilbert space with equal dimension for bosonic and fermionic states. Then the partition function will be
\beq
{\rm Tr} \, e^{-\beta H } = 2 \sum_j e^{-\beta \lambda_j}.
\eeq
Since the loop equations are normally derived in terms of sums over unequal eigenvalues, we need to multiply the answers from the $(\upalpha,\upbeta)=(1,2)$ ensemble by a factor of $2^{\text{ $\#$ of boundaries}}$. Just like we did for fermionic and unorientable JT gravity, keeping track of these factors are important for the duality to work.

\medskip

The partition function on the disk simply allows us to read of the spectral curve from the density of states 
\beq\label{eq:N1SC}
y(x)= - \frac{\cos(2\pi \sqrt{-x})}{\sqrt{-2x}}.
\eeq
We can verify now that the cylinder reproduces the matrix model. According to the matrix integral we should get from gravity four times the bosonic answer. Indeed each trumpet has an extra factor of $\sqrt{2}$ from the fermion path integral, while there is an extra $2$ from the sum over spin structures of the cylinder making a total factor of $4$.

\medskip

The loop equations for a general $(\upalpha,\upbeta)$ ensemble take the same form as in the previous lectures $2x y(x) R_{g,n+1}(x,I) + x F_{g,n}(x,I)\sim 0$ with
\bea
F_{g,n+1}(x,I) &=& \Big((1-\frac{2}{\upbeta}) \partial_x + \frac{\upalpha-1}{\upbeta x}\Big) R_{g-\frac{1}{2},n+1}(x,I)\nonumber\\
&&+ R_{g-1,n+2}(x,x,I) + \sum_{\rm stable} R_{h_1,|I_1|+1}(x,I_1) R_{h_2,|I_2|+1}(x,I_2)\nonumber\\
&&+ 2 \sum_{k=1}^n \Big( R_{0,2}(x,x_k) + \frac{1}{\upbeta} \frac{1}{(x-x_k)^2}\Big) R_{g,n}(x,I/x_k).\label{eq:LEAZ}
\ea
Curiously, when $(\upalpha,\upbeta)=(1,2)$ the form of the equations becomes precisely the same as that of the GUE ensemble. (This was in fact crucial for some proposals for non-perturbative completions of JT gravity by C. Johnson in \cite{Johnson:2019eik}.) This does not mean that the partition function of $\mathcal{N}=1$ supergravity is the same as its bosonic counterpart since the spectral curve in \eqref{eq:N1SC} is different.

An inverse Laplace transform of the loop equations leads to a similar recursion relation as the one that appeared in bosonic JT gravity in \eqref{eqn:MTRMTM} but with a kernel 
\beq\label{eq:DDDD}
D(x,y) = \int_{-\i \infty}^{\i \infty} \frac{\d z}{2\pi \i} \, \frac{e^{-x z}}{2z y(-z^2)} \, \sinh (yz)= \frac{1}{8\pi} \Big( \frac{1}{\cosh \frac{x-y}{4}} - \frac{1}{\cosh \frac{x+y}{4}} \Big).
\eeq
It was shown by Stanford and Witten that the Weil-Petersson volumes of $\mathcal{N}=1$ surfaces follow precisely a recursion relation with this kernel! This was also implied by an earlier work of Norbury \cite{Norbury_2023}. This proves that $\mathcal{N}=1$ JT gravity including the mod 2 index is dual to the $(\upalpha,\upbeta)=(1,2)$ matrix ensemble.

\medskip

If you recall the derivation of Mirzakhani recursion, it involves computing the length on a boundary circle of a segment where geodesics have certain properties. This singles out a three-holed sphere with no moduli. In the supergravity case the three-holed sphere has fermionic moduli and have to be integrated. This causes the changes in the recursion kernel that matches exactly with the loop equations.

\medskip
You might be confused since the theory with the $(-1)^\zeta$ is non-anomalous, while the theory without it is anomalous. This is the opposite as what we saw in fermionic gravity! The resolution is that the Schwarzian fermion contributes to the anomaly. The presence of a bulk $(-1)^\zeta$ cancels that anomaly, while it remains in the other theory.

\subsubsection{Unorientable $\mathcal{N}=1$ supergravity}

In the $BF$ language, in order to define unorientable supergravity we need to select an automorphism of the gauge group to identify with an orientation reversal. We will work with the basis of generators we wrote above $\{ {\sf e}, {\sf f}, {\sf h};{\sf q}_1 , {\sf q}_2 \}$. If we conjugate the bosonic generators by the orientation reversal ${\rm diag}(1,-1)$ we obtain 
\beq
{\sf e} \to - {\sf e},~~~~{\sf f} \to - {\sf f},~~~~{\sf h} \to {\sf h}.
\eeq
This preserves the ${\rm sl}(2,\mathbb{R})$ algebra $[{\sf h},{\sf e}]=2{\sf e}$, $[{\sf h},{\sf f}]=-{\sf f}$ and $[{\sf e},{\sf f}]={\sf h}$. What about fermions? We need to guarantee that the action of this transformation on fermions keeps the algebra unchanged. The relations ${\sf q}_1^2 = -{\sf e}$ and ${\sf q}_2^2 = {\sf f}$ imply that the transformation has to be ${\sf q}_1 \to \pm \i {\sf q}_1$ and ${\sf q}_2 \to \pm \i {\sf q}_2$. On the other hand $\{ {\sf q}_1,{\sf q}_2 \} = {\sf h}$ implies the two signs have to be opposite so if ${\sf q}_1\to \upeta \i {\sf q}_1$ then ${\sf q}_2 \to -\upeta \i {\sf q}_2$ where $\upeta=\pm 1$ is an arbitrary sign. The only other non-trivial commutators to check are $[{\sf e}, {\sf q}_2] = {\sf q}_1$ and $[{\sf f},{\sf q}_1]={\sf q}_2$. This is the only way to define an orientation reversal in the presence of one local supersymmetry.

\medskip 

Applying an orientation reversal twice (as defined in the previous paragraph) leaves the bosonic generators unchanged while it acts on fermions by reversing their sign. This means that such a transformation satisfies ${\sf R}^2 = (-1)^{\sf F}$, implying that $\mathcal{N}=1$ supergravity only allows for a sum over $\pin^-$ structures.

\medskip

One can provide a holographically dual argument for the restriction to $\pin^-$ structures. The CPT theorem implies that a $\pin^+$ structure leads to a time-reversal symmetry acting classically as ${\sf T}^2 = (-1)^{\sf F}$. This cannot happen for a theory with one supercharge; since the supercharge are fermionic they have to come in representations of ${\sf T}$ which are at least two-dimensional.

\medskip

We can therefore define 8 different unorientable $\mathcal{N}=1$ JT gravity by including the eta-invariant TQFT, which being topological is automatically supersymmetric. These theories saturate the remainder of the ten Altland-Zirnbauer random matrix ensembles. They all have measures of the form $(\upalpha,\upbeta)$ with different values of those parameters.

\subsection{$\mathcal{N}=2$ JT gravity}

The previous derivation was extended to $\mathcal{N}=2$ JT supergravity in \cite{Turiaci:2023jfa}. We will only have time to cover some main aspects of the calculation.

\subsubsection{Derivation of the matrix ensemble} 

In this case we will begin by analyzing the random matrix ensemble, since this has not been done for systems with extended supersymmetry before \cite{Turiaci:2023jfa}. 

\medskip

$\mathcal{N}=2$ quantum mechanics implies the existence of two charges $Q$ and $Q^\dagger$ that satisfy the algebra
\beq
Q^2 = Q^{\dagger}{}^2=0,~~~~\{ Q, Q^\dagger\} = H.
\eeq
We also assume the existence of an $R$-symetry $\U(1)$ generated by $J$ satisfying 
\beq
[J,Q] = Q ,~~~[J,H]=0.
\eeq
The operator distinguishing bosons and fermions is now given by 
\beq
(-1)^\F = e^{\pm \i \pi J}.
\eeq
The Hilbert space decomposes according to the spectrum of $R$-charges $\mathcal{H}=\oplus_k \mathcal{H}_k$. A state $\psi_k \in \mathcal{H}_k$ if $J\, \psi_k = k \, \psi_k$. The supercharge decomposes accordingly
\beq
Q = \sum_k Q_k ,~~~~Q_k : \mathcal{H}_k \to \mathcal{H}_{k+1},
\eeq
and the algebra has two types of irreducible multiplets. BPS multiplets are invariant under all supercharges $Q \psi_k = Q^\dagger \psi_k = 0$ and come as one state of charge $k$. Non-BPS multiplets come in pairs of charge $(k,k+1)$ and $Q\psi_k \sim \psi_{k+1}, Q^\dagger \psi_{k+1} \sim \psi_k$. The Hilbert space therefore further decomposes into
\beq
\mathcal{H}_k = \underbrace{\mathcal{H}_k^0}_{\text{BPS states of charge k}} \oplus \underbrace{\mathcal{H}_k^+}_{\text{From multiplet $(k,k+1)$}} \oplus \underbrace{\mathcal{H}_k^-}_{\text{From multiplet $(k,k-1)$}} 
\eeq
We denote by $L$ the total dimension of the Hilbert space, $L_k$ the dimension of $\mathcal{H}_k$ and by $L^0_k$ the number of BPS states of charge $k$. 

\medskip 

We would like to construct a random matrix model where we integrate over all supercharges satisfying the right algebra, and such that the ensemble is invariant under unitaries acting on each subsector of charge $k$. This leads to two immediate issues:

\begin{itemize}
    \item A naive prescription is to integrate over all $Q_k$ as if they were complex matrices. This is wrong since the algebra $Q^2=0$ imply that 
    \beq
    Q_k \cdot Q_{k-1} =0.
    \eeq
    This has to be supplemented as a constraint.

    \item Consider the symmetry group $\U(L_k)$ acting on $\mathcal{H}_k$. Both matrices $Q_{k-1}$ and $Q_k$ are  affected, via left- or right-multiplication, by this unitary transformation. Therefore in the reduction to its singular value integral, the measure one derives will not obviously factorize between different supermultiplets.
\end{itemize}
To say it in another words, we will compute the measure factor by reducing
\beq
\int \prod_{k} \, \d Q_k \, \prod_{s} \delta(Q_{s+1}\cdot Q_s) 
\eeq
to a suitable integral over ``eigenvalues''. This calculation was done in section 2 of \cite{Turiaci:2023jfa} with the result 
\begin{itemize}
    \item The energy spectrum of each $(k,k+1)$ multiplets are statistically independent of each other.
    \item Denoting by $\lambda_i$ the ``singular values'' of $Q_k$, the measure over each supermultiplet reduces to that of the Altland-Zirnbauer ensemble with 
    \beq
\prod_{i}\int \d \lambda_i\,|\lambda_i|^\upalpha \, \prod_{i<j} |\lambda_i^2-\lambda_j^2|^{\upbeta}~~~~\Rightarrow~~~(\upalpha, \upbeta) = (1+ 2 L_k^0+2 L^0_{k+1}, 2).
    \eeq
    In models where $L_k^0$ is proportional $L$ the BPS contribution to the measure can be absorbed in the matrix potential leading to effectively a $(1,2)$ model. Matrix models with logarithmic potentials are also called Penner models. In such cases the information of the BPS states is in the spectral curve, not in the loop equations.
    \item The wavefunction of supersymmetric states of $R$-charge $k$ inside $\mathcal{H}_k$ are random. This was the assumption behind the proposal by Lin, Maldacena, Rozenberg and Shan \cite{Lin:2022rzw,Lin:2022zxd} to detect chaos in the BPS sector of systems dual to black holes. See also \cite{Chen:2024oqv}.
\end{itemize}
The result is reasonable but yet surprising. It relies on a non-trivial cancellation between the effects enumerated above that want to correlate the singular values of different $Q_k$'s.

\subsubsection{Supergravity path integral}

In the $BF$ formalism we can define $\mathcal{N}=2$ JT gravity as a gauge theory with the supergroup $\SU(1,1|1)={\rm OSp}(2|2)/\mathbb{Z}_2$. (Another formulation in superspace is available \cite{Forste:2017apw} although not useful for our purposes.) This is the group of linear transformations acting on a space of dimension $2|1$ preserving an inner product. The maximal bosonic subgroup includes $\SL(2,\mathbb{R})$ and $\U(1)$. This implies that the theory includes bosonic JT gravity, a 2d Maxwell field, together with a complex gravitini and dilatini. The path integral localizes to flat $\SU(1,1|1)$ connections which in the appropriate component reduce to $\mathcal{N}=2$ hyperbolic surfaces. 

\medskip

The holonomy around a geodesic is conjugated to\footnote{One can also work with a $\widehat{q}$-fold cover of $\SU(1,1|1)$ where $\phi \sim \phi + 2\pi /\widehat{q}$ and $\widehat{q}$ is an odd integer. This generalization is important, particularly in the context of SYK \cite{Fu:2016vas}, but will not make any drastic change in the discussion here. There are also SYK-like models with $\hat{q}=1$ \cite{Heydeman:2022lse}.} 
\beq
U = e^{\i \phi } \left( \begin{array}{cc|c}
  - e^{b/2}   & 0 & 0  \\
    0  & - e^{-b/2} & 0 \\
    \hline 0 & 0 & e^{\i \phi}
\end{array}
\right).
\eeq
$b$ represents the geodesic length while $\phi$ labels the $\U(1)$ holonomy around it. We have fixed conventions where $\phi=0$ corresponds to antiperiodic fermions. Periodic fermions can be achieved by setting $\phi = \pi$, what would be referred to as spectral flow in the context of 2d CFT.

\medskip

A new feature of this theory is that already at the level of the disk topology we can insert topological theories in the bulk that affect the partition function. This is due to the fact that even at the disk level there is a sum over topologically inequivalent configurations of the $\U(1)$ gauge field labeled by their first Chern class $\int F$. Adding such a term is the analog of the theta angle in four dimensions. We call its coefficient $\delta$ and given that $\int F = \oint A$ it will appear in the dual quantum mechanical model as a background $\U(1)$ $R$-charge. A recent realization of this deformation in higher-dimensional near extremal black holes is in \cite{Heydeman:2024fgk,Heydeman:2024ezi}.

\medskip

NAdS boundaries are parametrized by the inverse temperature $\beta$ but also the chemical potential $\alpha$ which we define such that the partition function computes
\beq
Z(\beta,\alpha) = {\rm Tr}\, \left( e^{-\beta H} e^{\i \alpha J} \right).
\eeq
Since the $R$-charge spectrum is integral we expect $\alpha \sim \alpha + 2\pi$ although we will see this statement might present anomalies. One way to achieve the insertion of the chemical potential term is to impose twisted boundary conditions for charge fields $\psi(\tau + \beta) = \pm e^{\i \alpha J} \psi(\tau)$ around the thermal circle.

\medskip

The disk partition function reduces to an $\mathcal{N}=2$ version of the Schwarzian theory, which includes a $\U(1)$ phase mode and a complex fermion. The action still localizes and is one-loop exact. The final answer is 
\beq
Z_{\rm disk}(\beta,\alpha) = e^{S_0} \sum_{n\in\mathbb{Z}} \exp{(2\pi \i n \delta)} \, \frac{\cos(\frac{\alpha}{2}+\pi n)}{2\pi^3(1-4 (\alpha/2\pi + n)^2)} \, e^{\frac{\pi^2}{\beta}(1-4(\alpha/2\pi + n)^2}.
\eeq
The prefactor arises from the one-loop determinant. Since the fermions are now charged under the $\U(1)$ symmetry, the one-loop determinant depends non-trivially on the $\U(1)$ chemical potential. The sum over saddles involves different topological sectors of the $\U(1)$ field. We also included a weight $e^{2\pi \i n \delta}$ for each sector. Finally, there is no temperature dependence in the one-loop determinant since the isometries of the $\mathcal{N}=2$ hyperbolic disk have the same number of bosonic as fermionic generators.

\medskip

After a slightly involved inverse Laplace transform, the spectrum of the theory becomes
\bea
Z(\beta,\alpha) &=& \sum_{k \in \mathbb{Z}+\delta} e^{\i \alpha k} \, \underbrace{\frac{e^{S_0} \cos( \pi k)}{4\pi^2}}_{\text{BPS states of charge $k$, $L_k^0$}}\nonumber\\
&&\hspace{-1cm}+ \sum_{q\in \mathbb{Z}+\delta-\frac{1}{2}} (e^{\i \alpha( q-\frac{1}{2})} + e^{\i \alpha (q+\frac{1}{2})} ) \int_{E_0(q)}^\infty \d E \, e^{-\beta E} \underbrace{\frac{e^{S_0} \sinh (2\pi \sqrt{E-E_0(q)})}{8\pi^3 E}}_{\text{Spectrum of non-BPS multiplets}}.
\ea
where $E_0(q) = q^2/4$. The non-BPS multiplets have charges $(q-\frac{1}{2},q+\frac{1}{2})$. We see that the effect of the 2d theta angle is simply to produce a shift in the background charge of the dual quantum system by $\delta~{\rm mod}~\mathbb{Z}$. We find a large, meaning order $e^{S_0}$, number of BPS states of charge $k=\delta$ unless $\delta=1/2$. The non-BPS multiplets have a charge-dependent gap $E_0$. This is non-vanishing for all multiplets unless $\delta =1/2$ and $q=0$. We see the presence of a square-root edge for all multiplets with a gap, and an inverse-square-root edge for multiplet without a gap, consistent with the $(\upalpha,\upbeta)$ ensembles. For each multiplet this result implies the following spectral curve
\beq
y(x) = \frac{\sin(2\pi\sqrt{-x+q^2/4})}{8\pi^2 x}.
\eeq
with a threshold at $x=E_0(q)=q^2/4$. Some further aspects of this model were studied in\cite{Johnson:2023ofr} and the connection between the number of BPS states and the gap in each supermultiplet was studied in \cite{Johnson:2024tgg}.

\medskip

The cylinder path integral can be obtained from gluing two double-trumpets. The twist parameter comes now with a $\U(1)$ partner; the $\SU(1,1|1)$ holonomies describing parallel transport from one boundary to another, represented by matrices that commute with $U$ are proportional to ${\rm diag}(e^{\varrho/2},e^{-\varrho/2},-e^{\i \varphi})$. The trumpet partition function is 
\beq
Z_{\rm trumpet} = \sum_{n\in \mathbb{Z}} \exp{(2\pi \i n \delta)} \frac{\cos(\frac{\alpha}{2} + \pi n)}{\pi \beta} e^{-\frac{b^2}{4\beta} - \frac{4\pi^2}{\beta}(\frac{\alpha-\phi}{2\pi} + n)^2}.
\eeq
The final answer for the double-trumpet is
\bea
Z &=& (2\pi) \int_0^{2\pi} \d \phi \int_0^\infty b\, \d b Z_{\rm trumpet}(\beta_1,\alpha_1; b,\phi) Z_{\rm trumpet}(\beta_2,\alpha_2; b,-\phi)\nonumber\\
&&\hspace{-1cm}=\sum_{q} (e^{\i \alpha_1(q-\frac{1}{2})} + e^{\i \alpha_1(q+\frac{1}{2})})(e^{\i \alpha_2(q-\frac{1}{2})} + e^{\i \alpha_2(q+\frac{1}{2})}) \frac{\sqrt{\beta_1 \beta_2}}{2\pi (\beta_1+\beta_2)} e^{-\beta_1 E_0(q) - \beta_2 E_0(q)}.
\ea
This is consistent with the ensemble from \cite{Turiaci:2023jfa} which predicts that the supermultiplets should be statistically independent; notice that there is a single sum over multiplets $q$ instead of two. This is achieved concretely by the integral over intermediate $\U(1)$ holonomies along the internal circle projecting into contributions where the same multiplet propagates in both boundaries.

\medskip

More complicated surfaces can again be built out of three-holed spheres glued to trumpets. The obvious next step is to evaluate the torsion. At this point, one can apply the same approach we have done so far to the group $\SU(1,1|1)$. This is easier said than done, and multiple subtleties arise that need to be taken care of. These can be read in \cite{Turiaci:2023jfa}. Instead, we will just point out some salient features of some results.

\medskip

For example, in $\mathcal{N}=1$ supergravity we argued that all genus zero volumes vanish due to the presence of fermionic moduli. Is this the case in $\mathcal{N}=2$ supergravity? The answer is no. The path integral on the three-holed sphere for example is given by
\beq
V_{0,3} = - \frac{1}{2\pi} \frac{1}{4} \delta''(\phi_1+\phi_2+\phi_3).
\eeq
The first feature we see is the presence of the delta function imposing $\phi_1+\phi_2+\phi_3=0$. This can be understood directly from considering the $\U(1)$ gauge field; there are no flat connections unless this constraint is satisfied. The constraint is still satisfied in supergravity but it gets fermionic corrections where now 
\beq
\phi_1 + \phi_2 + \phi_3 = (\text{fermions}).
\eeq
The precise form of the fermion terms can be derived from the condition $UVW=1$ on the $\SU(1,1|1)$ holonomies. Now when we integrate over the fermionic moduli, non-vanishing terms can be picked up from the fermionic terms of the constraint. To give an example consider the integral over two odd variables $\theta_1$ and $\theta_2$
\bea
\int \d^2 \theta \, \delta( x+ \theta_1 \theta_2)&=&\int \d^2 \theta \,\left(\delta(x) + \delta' (x) \theta_1\theta_2 \right),\nonumber\\
&=& \delta'(x)
\ea
This type of effect leads to the derivatives acting on the delta function.  

\medskip

It is convenient to present these volumes in terms of their Fourier transforms (to ease the notation we leave implicit that a volume with $n$ boundaries comes with $\{b_1,\ldots,b_n\}$ arguments)
\beq
V_{g,n}(q) = (2\pi)^{2-2g} \int \frac{\d^n \phi}{(2\pi)^n} e^{\i q \sum_j \phi_j} V_{g,n}.
\eeq
One can show that these volumes are polynomials in both $b$ and $q$ 
\beq
V_{g,n}(q) = \sum_{m=1}^{2g-2+n} (q^2/4)^{m} v_{g,n,m}(b_1,\ldots,b_n).
\eeq
There are a few properties one can easily show by deriving a supersymmetric generalization of Mirzakhani recursion: 

\begin{itemize} 

\item The volumes are polynomials in $q^2$ of highest degree equal to $2g-2+n$.

\item The term with the highest power of $q$ is equal to the bosonic volumes computed by Mirzakhani.

\item As we decrease the power of $q^2$ by $s$ units, the coefficient $v_{g,n,m=2g-2+n-s}$ is a polynomial in $b^2$ of degree $3g-3+n-s$. In particular $v_{g,n,0}$ is a polynomial of degree $g-1$. This is also true for $g=0$ since $v_{0,n,0}=0$ vanishes.

\item Some examples one can easily evaluate from the recursion:
\bea
V_{0,3} &=& \frac{q^2}{4} 1 ,~~~V_{1,1} = \frac{q^2}{4} \frac{b^2 + 4\pi^2}{48} + \Big(-\frac{1}{8}\Big),\nonumber\\
V_{0,4}&=& \frac{q^4}{16} \frac{4\pi^2 + b_1^2+b_2^2+b_3^2+b_4^2}{2} + \frac{q^2}{4} (-3).
\ea
\end{itemize}

We can repeat the derivation of Mirzakhani taking into account the fermionic and bosonic extra moduli involved in the three-holed sphere that is used to apply her argument. This produces yet a new set of kernels and applying a Laplace transform similar to what we did in the second lecture, one can show that the GPI of $\mathcal{N}=2$ JT supergravity is dual to the random matrix ensemble described earlier.

\medskip 

Finally, we can consider theories with time-reversal symmetry. One can define two classes of theories depending on whether ${\sf T}$ commutes or anti-commutes with $J$. For each case there are two further choices depending on the sign of ${\sf T}^2$. In \cite{Turiaci:2023jfa} these cases were considered, shown to reduce to statistically independent Altland-Zirnbauer ensembles, and related to different theories of unorientable $\mathcal{N}=2$ supergravity, although a full proof of the duality was not completed. One could consider a case with separate charge conjugation ${\sf C}$ and time-reversal ${\sf T}$ symmetries which has not been studied so far. Another interesting generalization would be to theories where supersymmetry is softly broken \cite{Heydeman:2024ohc}.

\section{Conclusions}\label{conclusions}

In these lectures, we covered some technical aspects regarding the evaluation of the GPI for simple models of 2d gravity, and their dualities with random matrix models. To conclude, here is a list of some recent research directions:

\paragraph{JT gravity with end-of-the-world branes:} An interesting generalization involves the addition of end-of-the-world branes to JT gravity. This allows us to use the GPI to prepare pure states in gravity. This was derived by Penington, Shenker, Stanford and Yang \cite{Penington:2019kki} in the context of introducing replica wormholes to show that the dimension of the Hilbert space of black hole microstates is finite, showing a drastic reduction from a semiclassical analysis around the black hole background.
 
 \paragraph{JT gravity with propagating matter:} It would be valuable to generalize the results here to theories of gravity coupled to matter. One motivation to do this is the fact that top-down models of JT gravity involve dilaton-gravity coupled to matter. Another motivation is to unify a picture of quantum chaos in the sense presented here to some aspects of thermalization, such as the eigenstate thermalization hypothesis and its relation to spacetime wormholes. Recent proposals in this direction are \cite{Jafferis:2022uhu,Jafferis:2022wez}. The main obstruction here is the fact that in the presence of unitary matter fields the GPI on spacetime wormholes diverges. Whether this divergence is physical or an unwanted feature of simplified toy models is yet unclear.

\paragraph{Reproducing the plateau from gravity:} The spectral form factor of a chaotic system consists of a dip, reproduced by the black hole geometry, a ramp, reproduced by the double-cone \cite{Saad:2018bqo}, and a late-time plateau. A satisfactory picture in gravity for the plateau is missing. Some recent proposals are \cite{Johnson:2020exp} or \cite{Blommaert:2022lbh,Saad:2022kfe} although they seem to rely heavily on the details of JT gravity, as opposed to for example the ramp which is universal.

\paragraph{Resolutions of factorization puzzle:} Pure dilaton-gravity is dual to an ensemble of quantum systems. In physical systems we expect wormholes to be responsible for the features characteristic of quantum chaos, even if the systems are not ultimately an ensemble. What is the physical principle behind the restoration of factorization is an open problems. Several proposals in the context of 2d gravity are available \cite{Blommaert:2019wfy, Blommaert:2021gha,Blommaert:2021fob} although the ultimate mechanism in top-down models such as in string theory and AdS/CFT is still unclear.

\paragraph{Three-dimensional gravity:} As briefly mentioned above, the developments in 2d gravity and its random matrix dual motivated a proposal to fix some pathologies in the GPI of pure 3d gravity \cite{Maxfield:2020ale}. This made it reasonable to expect an ${\rm AdS}_3/{\rm CFT}_2$ version of the SSS duality, a point made in \cite{Maxfield:2020ale} and also independently in \cite{Cotler:2020ugk}. 

 \paragraph{Firewalls:} An interesting question is whether spacetime wormholes can have an effect on the experience of an observer who falls into an old enough black hole. This question was studied by Stanford and Yang in \cite{Stanford:2022fdt}, where they suggested that wormhole effects can produce firewalls at the horizon of old black holes. This question was also further studied in\cite{Blommaert:2024ftn,Iliesiu:2024cnh} and remains an open problem.

\bigskip 

An interesting topic that is part of C. Johnson's lectures is the connection between the dualities of dilaton-gravity and random matrices with similar dualities for the minimal string theory \cite{Saad:2019lba,Johnson:2019eik,Mertens:2020hbs,Turiaci:2020fjj}. New versions of these dualities were recently proposed in \cite{Collier:2023cyw, Collier:2024kmo}. Finally, a generalization of the SSS duality to asymptotically de Sitter spacetimes was first proposed in \cite{Maldacena:2019cbz} and \cite{Cotler:2019nbi} and it is still an active area of investigation.

\bigskip

\textit{Acknowledgments} It is a pleasure to thank the participants of the 2024 Les Houches school on Quantum Geometry, as well as the organizers, and Clifford Johnson for useful conversations while preparing the lectures and for comments on the draft. I would also like to thank Thomas Mertens for discussions while preparing the review for Living Reviews in Relativity, Douglas Stanford for discussions while preparing a review talk for Strings 2024, and Chih-Hung Wu for carefully reading a draft. I was supported by the University of Washington and the DOE award DE-SC0024363.

\bibliographystyle{utphys2}
{\small \bibliography{Biblio}{}}

\end{document}